\documentclass[12pt]{article}
\usepackage{fullpage,citesort,
epsfig,psfrag,graphics,amsbsy,amssymb,amsmath,
}
\usepackage{caption}
\newcommand{\beq}{\begin{equation}}
\newcommand{\eeq}{\end{equation}}
\newcommand{\bea}{\begin{eqnarray}}
\newcommand{\eea}{\end{eqnarray}}
\newcommand{\bfs}{\boldsymbol}

\newcommand{\Tr}{{\rm Tr}}

\newcommand{\ket}[1]{|#1\rangle}
\newcommand{\bra}[1]{\langle#1|}

\def\math{\mathsurround=0pt }
\def\leftrightarrowfill{$\math \mathord\leftarrow \mkern-6mu
 \cleaders\hbox{$\mkern-2mu \mathord- \mkern-2mu$}\hfill
 \mkern-6mu \mathord\rightarrow$}
\def\overleftrightarrow#1{\vbox{\ialign{##\crcr
     \leftrightarrowfill\crcr\noalign{\kern-1pt\nointerlineskip}
     $\hfil\displaystyle{#1}\hfil$\crcr}}}

\newcommand{\VEV}[1]{\left\langle#1\right\rangle}

\let\l=\lambda

 \def\bd{\begin{document}} \def\ed{\end{document}}
\def\ds{\documentstyle} \let\fr=\frac \let\bl=\bigl \let\br=\bigr
\let\Br=\Bigr \let\Bl=\Bigl
\let\bm=\bibitem
\let\na=\nabla
\let\pa=\partial \let\ov=\overline
\def\ft#1#2{{\textstyle{{\scriptstyle #1}\over {\scriptstyle #2}}}}
\def\fft#1#2{{#1 \over #2}}
\def\vp{\varphi}
\def\sst#1{{\scriptscriptstyle #1}}
\def\oneone{\rlap 1\mkern4mu{\rm l}}
\def\td{\tilde}
\def\wtd{\widetilde}
\def\dalemb#1#2{{\vbox{\hrule height .#2pt
        \hbox{\vrule width.#2pt height#1pt \kern#1pt
                \vrule width.#2pt}
        \hrule height.#2pt}}}
\def\square{\mathord{\dalemb{6.8}{7}\hbox{\hskip1pt}}}
\def\wtd{\widetilde}
\def\R{\rlap{\rm I}\mkern3mu{\rm R}}
\def\im{{\rm i}}
\def\tilg{\tilde{g}}
\def\tilF{\tilde{F}}
\def\tilA{\tilde{A}}
\def\varf{\varphi}
\def\tilf{\tilde{\phi}}
\def\tilh{\tilde{h}}
\def\rme{{\rm e}}
\def\ep{\epsilon}
\def\0{{(0)}}
\def\9{{(9)}}
\def\8{{(8)}}
\def\7{{(7)}}
\def\6{{(6)}}
\def\5{{(5)}}
\def\4{{(4)}}
\def\3{{(3)}}
\def\2{{(2)}}
\def\1{{(1)}}
\newcommand{\trace}{{\rm Tr}}
\newcommand{\ub}{\overline{U}}
\newcommand{\vb}{\overline{V}}
\newcommand{\uh}{\widehat{U}}
\newcommand{\vh}{\widehat{V}}
\newcommand{\ubh}{\overline{\widehat{U}}}
\newcommand{\vbh}{\overline{\widehat{V}}}
\newcommand{\lb}{\bar{\l}}
\newcommand{\Fb}{\overline{F}}
\newcommand{\Fh}{\widehat{F}}
\newcommand{\Fbh}{\overline{\widehat{F}}}
\newcommand{\Ab}{\overline{A}}
\newcommand{\Ah}{\widehat{A}}
\newcommand{\Abh}{\overline{\widehat{A}}}
\newcommand{\Gb}{\overline{G}}
\newcommand{\Gh}{\widehat{G}}
\newcommand{\Gbh}{\overline{\widehat{G}}}
\newcommand{\Pb}{\overline{P}}
\newcommand{\Ph}{\widehat{P}}
\newcommand{\Pbh}{\overline{\widehat{P}}}
\newcommand{\Qb}{\overline{Q}}
\newcommand{\Qh}{\widehat{Q}}
\newcommand{\Qbh}{\overline{\widehat{Q}}}
\newcommand{\Bb}{\overline{B}}
\newcommand{\Bh}{\widehat{B}}
\newcommand{\Bbh}{\overline{\widehat{B}}}
\newcommand{\fhns}{\hat{F}^{\rm (NS)}}
\newcommand{\fhrr}{\hat{F}^{\rm (RR)}}
\newcommand{\ahns}{\hat{A}^{\rm (NS)}}
\newcommand{\ahrr}{\hat{A}^{\rm (RR)}}
\newcommand{\hhrr}{\hat{H}^{\rm (RR)}}
\newcommand{\hchi}{\hat{\chi}}
\newcommand{\hphi}{\hat{\phi}}
\newcommand{\htau}{\hat{\tau}}
\newcommand{\cG}{{\cal G}}
\newcommand{\cGb}{\overline{{\cal G}}}
\newcommand{\cH}{{\cal H}}
\newcommand{\cP}{{\cal P}}
\newcommand{\cPb}{\overline{{\cal P}}}
\newcommand{\cQ}{{\cal Q}}
\newcommand{\cQb}{\overline{{\cal Q}}}
\newcommand{\cM}{{\cal M}}
\newcommand{\cN}{{\cal N}}
\newcommand{\cO}{{\cal O}}
\newcommand{\cD}{{\cal D}}
\newcommand{\cL}{{\cal L}}
\newcommand{\vpp}{\mbox{$\langle{\scriptstyle++}\rangle$}}
\newcommand{\vmp}{\mbox{$\langle{\scriptstyle-+}\rangle$}}
\newcommand{\vppp}{\mbox{$\langle{\scriptstyle+++}\rangle$}}
\newcommand{\vmpp}{\mbox{$\langle{\scriptstyle-++}\rangle$}}
\newcommand{\vpmp}{\mbox{$\langle{\scriptstyle+-+}\rangle$}}
\begin{document}
\setlength{\captionmargin}{36pt}
\setlength{\captionmargin}{36pt}
\begin{titlepage}
\begin{flushright}
  \phantom{UFIFT-HEP}
\end{flushright}
\vskip 3cm
\begin{center}
\begin{large}
{\bf String Bit Description of Antiperiodic\\ Fermion  Worldsheet  Fields}
\end{large}

\vskip 2cm
{\large
 Charles B. Thorn\footnote{E-mail  address: {\tt thorn@phys.ufl.edu}}
}
\vskip0.20cm
{\it Institute for Fundamental Theory,\\
Department of Physics, University of Florida,
Gainesville FL 32611}


\vskip 1.0cm
\end{center}

\begin{abstract}
\noindent  We study a string bit Hamiltonian whose continuum limit describes
  antiperiodic (AP) anticommuting worldsheet fields. We calculate the amplitude
  for transitions between an AP spin chain and a periodic (P) one
  in the continuum limit, $M\to\infty$ where $M$ is the bit number of either chain.
  We also numerically evaluate the corresponding
  amplitudes at increasing finite $M$ to assess the convergence rate to the continuum.
  We then give the overlap equations for the transition AP$+$AP$\to$AP,
  and numerically solve them
  for increasing $M$ values at a fixed value of $x=K/M$,
  where $M$ is the bit number of the large chain and $K$ is the bit number of one
  of the smaller chains. For this case,
  in contrast to the situation with an even number of AP chains, there is an obstacle to directly finding the continuum limit analytically. We suggest
  an indirect analytic approach to this problem: using the AP$\to$P transition
  followed by a P$\to$AP$+$AP transition,
  each of which has a relatively simple analytic continuum limit.
  We also show how bosonization of the fermion fields enables an
  analytic recursive evaluation of the  AP$+$AP$\to$AP amplitude.
\end{abstract}
\vfill
\end{titlepage}
\section{Introduction}
The string bit concept \cite{thornsakh}
provides a vehicle for the holographic emergence \cite{thoofthologram} of
space from a quantum mechanical system of bits, where only a finite number of
states are available to each bit \cite{sunthorn,thornspace}. Each bit
creation operator is an $N\times N$ matrix
$(\phi_{a_1\cdots a_n}^\dagger)_\alpha^\beta$.
The Hamiltonian for the underlying
quantum system is assumed to be a single trace operator which commutes with
bit number $M=\Tr \sum \phi^\dagger\phi$.
Then the 't Hooft limit $N\to\infty$ \cite{thooftlargen}
implies that the energy eigenstates are
a collection of noninteracting string bit chains,
which for very large bit number behave as continuous strings.
In the $1/N$ expansion one can represent a closed string as a linear
combination of single trace states of the form \cite{thornprotobits}
\bea
\Tr \phi^\dagger_1\cdots \phi^\dagger_n\ket{0}.
\eea
where $\phi\ket{0}=0$ for all $\phi$. 
Then the string bit Hamiltonian applied to such a linear combination
behaves at large $N$ as
\bea
H\ket{\Psi_0}=h\ket{\Psi_0} +O(N^{-1})\times {\rm multi~trace~ states}
\eea
where $h$, which maps single trace states to single-trace states, acts
as the effective Hamiltonian that describes the closed chain at zeroth
order in the $1/N$ expansion \footnote{This $h$ is analogous, in quantum
  field theory, to  the   ``first quantized''
  Hamiltonian $h=\sqrt{{\bfs p}^2+m^2}$ of a free single
    particle at zero coupling. In this article we shall use
    upper case $H$ to denote the full
    (``second-quantized'') Hamiltonian and lower case $h$
    the corresponding first quantized Hamiltonian obtained, for
    string bit models when $N\to\infty$.} .

In the original formulation of string bit models,
\cite{thornsakh}, in addition to discrete labels, $\phi$ also depended on
$d=D-2$ continuous coordinates ${\bfs x}$, which label
points in the transverse space
in lightcone quantization of a string moving in $D-1$ dimensional space
\cite{goddardgrt,mandelstamlc,goldstone}.
The proposal of \cite{sunthorn,thornspace} is that each transverse coordinate
can emerge from a spin system. In
\cite{thornheis} we explored this mechanism in great
detail for the Heisenberg spin chain
 \bea
h&=&\frac{1}{4}\sum_{k=1}^M(\sigma_k^x\sigma_{k+1}^x+\sigma_k^y\sigma_{k+1}^y
+\Delta\sigma_k^z\sigma_{k+1}^z)
\eea
at the free fermion point $\Delta=0$. The Jordan-Wigner trick converts
$\sigma^{x,y}_k$ into spin variables $S_k^{x,y}$
that anticommute at different sites.
However then the resulting Fermi fields are periodic or antiperiodic
depending on the values of $M$ and $Q=\sum_k\sigma^z$, which commute with the
Hamiltonian. That is,
some sectors are
described by periodic fields while in other sectors by antiperiodic fields. It
turned out that the overlap amplitude between a large chain and two
smaller chains was nonzero in given sectors only if the three chains
involved included an even number of periodic sectors and an odd number of
antiperiodic sectors--a pattern that matches the three string vertices of
the RNS (Ramond-Neveu-Schwarz) \cite{rns} string theory and the
corresponding formulation of the superstring \cite{gso}.

Oddly this pattern is the reverse of the GS (Green-Schwarz)
superstring \cite{greenschwarz}. In the GS case, closed form
formulas for the interaction vertices can be found by various methods
\cite{greenschwarzbrink}.
In contrast the RNS pattern of vertices frustrates the known direct methods
for obtaining the vertices in closed form. An indirect method.
based on factorizing RNS dual resonance amplitudes
on onshell DDF states in the critical dimension, was used by
Hornfeck \cite{hornfeck,berkovits} to obtain an expression
for the vertex as a double sum over expressions similar to the vertices for the
bosonic string or GS superstring. This procedure gives the full onshell
RNS vertex in the critical dimension,
including the necessary operator prefactor.  To dig out the contribution of
the overlap discussed in this article would require further work.

\begin{figure}[ht]
\begin{center}
\includegraphics[width=3in]{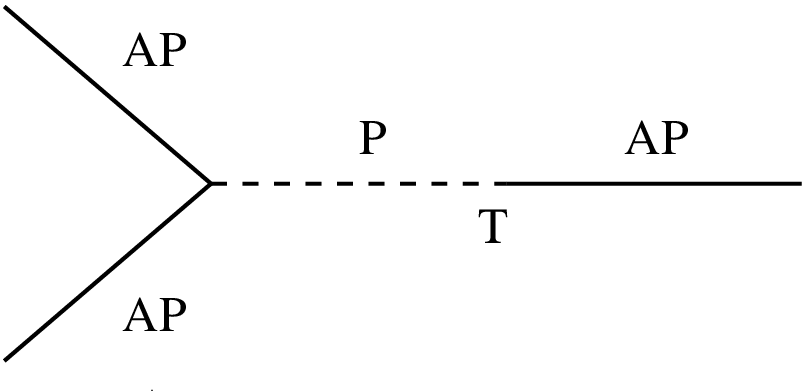}
\caption{Construction of vertex for 3 antiperiodic (AP)
  chains from the vertices for two antiperiodic chains
  to one periodic (P) chain and 1 periodic chain to one
antiperiodic chain. The desired vertex is the $T\to0$ limit.}
\label{apappap}
\end{center}
\end{figure}

In the present article we study
the two string transition between a periodic sector and an antiperiodic sector
which can be evaluated explicitly. With this transition amplitude in
hand one can hope to shed light on the relation of the two formulations
of superstring theory to each other. Specifically we have in mind 
another indirect construction of the vertex
for 3 antiperiodic chains as illustrated in
Fig.~\ref{apappap}. The desired vertex would then be obtained by
taking the $T\to0$ limit {\it after} the continuum limit. 
The two vertices
in this diagram will be obtained in this article, but we shall not attempt
the implied sum over $P$ states.

In section 2 we give our choice for the ``first quantized''
string bit Hamiltonian $h_{AP}$ which implements antiperiodic boundary
conditions on the worldsheet effective fermion fields, We obtain its
eigenvalue spectrum and evaluate some simple expectation values.
In section 3
we formulate the $AP\to P$ transition amplitude at finite $M$ and then
obtain its continuum limit $M\to\infty$. In section 4 we formulate the
finite $M$ three chain transition $AP\to AP+AP$, and discuss the obstacles
to finding its continuum limit analytically. Numerical calculations of
selected transitions for finite $M$ values ranging from
$M=14$ to $M=1792$ show convergence to a limiting value as $M$ gets large.
We also describe in a few cases how this limiting value can be analytically
obtained using the hypothesis of bosonization.
In Appendices we collect some known results for the $P\to P+P$ transition
amplitude and calculate the continuum limit of the $P\to AP+AP$ transitions
which could be converted to the $AP\to AP+AP$ by applying the $AP\to P$
transition obtained in this article to the $P$ leg.

\section{String bits for antiperiodic fermion worldsheet fields}
We use the same notation established in \cite{thornprotobits,songge}, which
dealt with spinor fields $S^a,{\tilde S}^a$ which obeyed periodic
boundary conditions. The index $a$ labels several independent spinor fields.
 This $P$ boundary condition
was implemented at the discrete level by substituting
$S^a_{M+1}\to S^a_1$, ${\tilde S}^a_{M+1}\to {\tilde S}^a_1$
in the expression for the Hamiltonian $h$.
This choice was actually dictated by our chosen string bit Hamiltonian;
which was designed to describe the superstring.
Introducing a minus sign in this substitution $S^a_{M+1}\to -S^a_1$,
${\tilde S}^a_{M+1}
\to -{\tilde S}^a_1$ leads to antiperiodic (AP) boundary conditions 
a la Neveu-Schwarz. The resulting contribution to $h$ is then
\bea
h_{AP}&=&
\sum_{k=1}^{M-1}\left[-iS^a_kS^a_{k+1}
  +i{\widetilde S}^a_k{\widetilde S}^a_{k+1}-iS_k^a{\widetilde S}^a_{k+1}
  +i{\widetilde S^a}_kS^a_{k+1}
+2iS^a_k{\widetilde S}^a_k\right]\nonumber\\
&&-\left[-iS^a_MS^a_{1}+i{\widetilde S}^a_M{\widetilde S}^a_{1}
-iS_M^a{\widetilde S}^a_{1}+i{\widetilde S^a}_MS^a_{1}
-2iS^a_M{\widetilde S}^a_M\right].
\eea
Since $h$ is bilinear in the spin variables, its eigenvalues can be found by
finding energy raising and lowering operators.
To find them we compute, for $k=2,\ldots,M-1$, 
\bea
{}[h_{AP},S_k^a]&=&2i(S_{k+1}-S_{k-1}+{\widetilde S}^a_{k+1}
+{\widetilde S}^a_{k-1}-2{\widetilde S}^a_{k})\\
{}[h_{AP},{\widetilde S}_k^a]&=&-2i({\widetilde S}_{k+1}-{\widetilde S}_{k-1}
+{S}^a_{k+1}
+{S}^a_{k-1}-2{ S}^a_{k})
\eea
 We then handle $k=1,M$ separately
\bea
{}[h_{AP},S_1^a]&=&2i(S_{2}+S_{M}+{\widetilde S}^a_{2}
-{\widetilde S}^a_{M}-2{\widetilde S}^a_{1})\\
{}[h_{AP},{\widetilde S}_1^a]&=&-2i({\widetilde S}_{2}+{\widetilde S}_{M}
+{S}^a_{2}
-{S}^a_{M}-2{ S}^a_{1})\\
{}[h_{AP},S_M^a]&=&2i(-S_{1}-S_{M-1}-{\widetilde S}^a_{1}
+{\widetilde S}^a_{M-1}-2{\widetilde S}^a_{M})\\
{}[h_{AP},{\widetilde S}_M^a]&=&-2i(-{\widetilde S}_{1}-{\widetilde S}_{M-1}
-{S}^a_{1}
+{S}^a_{M-1}-2{ S}^a_{M})
\eea
In the rest of this section we suppress the label $a$ to reduce clutter.
We see that these special cases can be incorporated into the general $k$
equations with the identifications, suppressing the $a$ label,
$S_0=-S_M$, $S_{M+1}=-S_1$, 
${\widetilde S}_0=-{\widetilde S}_M$, ${\widetilde S}_{M+1}=-{\widetilde S}_1$.
This means that in solving the eigenoperator equations by Fourier transforms,
we have to use half odd integer modes:
\bea
S_k&=&\frac{1}{\sqrt{M}}\sum_{r=1/2}^{M-1/2}e^{2\pi ikr/M}D_r,\qquad
{\widetilde S}_k=\frac{1}{\sqrt{M}}\sum_{r=1/2}^{M-1/2}e^{2\pi ikr/M}
{\widetilde D}_r\\
D_r&=&\frac{1}{\sqrt{M}}\sum_{k=1}^{M}e^{-2\pi ikr/M}S_k,\qquad
{\widetilde D}_r=\frac{1}{\sqrt{M}}\sum_{k=1}^{M}e^{-2\pi ikr/M}
{\widetilde S}_k\eea
Then
\bea
{}[h_{AP},D_r]&=&2i(e^{2\pi i r/M}-e^{-2\pi i r/M})D_r
+2i(e^{2\pi i r/M}+e^{-2\pi i r/M}-2){\widetilde D}_{r}\nonumber\\
&=&
-8\sin\frac{\pi r}{M}\left[D_r\cos\frac{\pi r}{M}+i{\widetilde D}_{r}
\sin\frac{\pi r}{M}\right]
\\
{}[h_{AP},{\widetilde D}_r]&=&8\sin\frac{\pi r}{M}\left[{\widetilde D}_r\cos\frac{\pi r}{M}+i{D}_{r}
\sin\frac{\pi r}{M}\right]
\eea
Then we look for eigenoperators
\bea
{}[h_{AP},D_r+\xi{\widetilde D}_r]&=&8\sin\frac{\pi r}{M}
\left[{\widetilde D}_r\left(\xi\cos\frac{\pi r}{M}
    -i\sin\frac{\pi r}{M}\right)+{D}_{r}
\left(i\xi\sin\frac{\pi r}{M}-\cos\frac{\pi r}{M}\right)\right]\nonumber\\
&\equiv&\Delta(D_r+\xi{\widetilde D}_r)
\eea
The analysis is identical to the periodic case:
\bea
\xi_\pm&=&\begin{cases}\displaystyle{i\tan\frac{\pi r}{2M}}\\
\phantom{.}\\
\displaystyle{-i\cot\frac{\pi r}{2M}}\end{cases}
\nonumber\\
\Delta_\pm&=&\mp8\sin\frac{\pi r}{M}.
\eea
The energy lowering operators are
\bea
G_r&=& D_r\cos\frac{\pi r}{2M}+i{\widetilde D}_r\sin\frac{\pi r}{2M},
\eea
and the raising operators are
\bea
{\bar G}_r&=&D_r\sin\frac{\pi r}{2M}-i{\widetilde D}_r\cos\frac{\pi r}{2M},
\eea
which can be inverted
\bea
D_r&=&G_r\cos\frac{\pi r}{2M}+{\bar G}_r\sin\frac{\pi r}{2M},\qquad
i{\widetilde D}_r=G_r\sin\frac{\pi r}{2M}-{\bar G}_r\cos\frac{\pi r}{2M}.
\eea
We notice that
\bea
G_r^\dagger&=&D_{M-r}\cos\frac{\pi r}{2M}-i{\widetilde D}_{M-r}\sin\frac{\pi r}{2M}
={\bar G}_{M-r}\\
\{G_r,{\bar G}_s\}&=&2\sin\frac{\pi r}{2M}\cos\frac{\pi s}{2M}\delta_{r+s,M}
+2\cos\frac{\pi r}{2M}\sin\frac{\pi s}{2M}\delta_{r+s,M}=2\delta_{r+s,M}\\
\{G_r,{G}^\dagger_s\}&=&2\delta_{rs}
\eea
We next express $h_{AP}$ in terms of the raising and lowering operators
\bea
h_{AP}&=&2\sum_{r=1/2}^{M-1/2}\sin\frac{r\pi}{M}\left[G_r^\dagger G_r-G_r 
G_r^\dagger\right]=2\sum_{r=1/2}^{M-1/2}\sin\frac{r\pi}{M}\left[2G_r^\dagger G_r-2\right]
\eea
From which we read off the ground state energy, for $s$ independent
spinor fields
\bea
E_G^{AP}=-4s\sum_{r=1/2}^{M-1/2}\sin\frac{r\pi}{M}=-\frac{4s}{\sin(\pi/(2M))}
\sim -\frac{8s}{\pi}M-\frac{\pi s}{3M}+O(M^{-3})
\eea
\subsection{Expectation values}
A nice exercise is to calculate the expectation value of $h_{AP}$
in the ground state of $h_P$ or the expectation of $h_P$ in the ground state
of $h_{AP}$. To do this we use
\bea
h_{AP}&=&h_P-2\left[-iS_MS_1+i{\widetilde S}_M{\widetilde S}_1
-iS_M{\widetilde S}_1+i{\widetilde S}_MS_1\right]
\eea
The terms in square brackets can be expanded either in terms of periodic
modes to calculate $\VEV{G_P|h_{AP}|G_P}$ or in terms of
antiperiodic modes to calculate $\VEV{G_{AP}|h_{P}|G_{AP}}$.
Starting with the first one,
\bea
\VEV{G_P|h_{AP}|G_P}&=&E_G^P-\frac{2}{M}\sum_{m,n=0}^{M-1}e^{2\pi in/M}
i\VEV{G_P|\left[-B_mB_n+{\widetilde B}_m{\widetilde B}_n
-B_m{\widetilde B}_n+{\widetilde B}_mB_n\right]|G_P}\nonumber\\
&=&E_G^P-\frac{2}{M}\sum_{m,n=0}^{M-1}e^{2\pi in/M}
i\VEV{G_P\left|\left[({\widetilde B}_m-B_m)(B_n+{\widetilde B}_n)
\right]\right|G_P}
\eea
So we need
\bea
\VEV{G_P\left|\left[({\widetilde B}_0-B_0)(B_0+{\widetilde B}_0)
\right]\right|G_P}=\VEV{G_P\left|2{\widetilde B}_0B_0
\right|G_P}
\eea
for zero modes, and for nonzero modes
\bea
\VEV{G_P\left|({\widetilde B}_m-B_m)(B_n+{\widetilde B}_n)
\right|G_P}&=&-\VEV{G_P\left|(F_m-i{\bar F}_m)(F_n+i{\bar F}_n)
\right|G_P}e^{i(m-n)\pi/2M}\nonumber\\
&=&-\VEV{G_P\left|iF_m{\bar F}_n)\right|G_P}e^{i(m-n)\pi/2M}\nonumber\\
&=&2\delta_{m+n,M}e^{-in\pi/M}
\eea
Then
\bea
\VEV{G_P|h_{AP}|G_P}&=&E_G^S-\frac{4}{M}\VEV{G_P\left|i{\widetilde B}_0B_0
\right|G_P}-\frac{4i}{M}\sum_{n=1}^{M-1}e^{i\pi n/M}\nonumber\\
&=&E_G^S-\frac{4}{M}\VEV{G_P\left|i{\widetilde B}_0B_0
\right|G_P}+\frac{4}{M}\cot\frac{\pi}{2M}
\eea
The square of the operator $i{\widetilde B}_0B_0$ is $1$ implying that its
eigenvalues are $\pm1$.
It is noteworthy that $\VEV{G_P|h_{AP}|G_P}-E^S_G\to8/\pi=O(1)$ at large
bit number, which is much larger than the excitation energies $=O(1/M)$
in the same limit. 

One can repeat the exercise for the expectation value of $h_P$ in
the ground state of $h_{AP}$. In this case there are no zero modes
and we have
\bea
\VEV{G_{AP}|h_{P}|G_{AP}}
&=&E_G^{AP}-\frac{4i}{M}\sum_{r=1/2}^{M-1/2}e^{i\pi r/M}\nonumber\\
&=&E_G^{AP}+\frac{4}{M}\csc\frac{\pi}{2M}
\eea
In each case the expectation of the ``wrong'' $h$ is much higher than the
ground energy of the ``right'' $h$, consistently
with the variational principle.

\section{Periodic/Antiperiodic Transition Amplitude}
A very simple overlap problem is the relationship of eigenstates 
of a Hamiltonian with periodic boundary conditions to those of
a Hamiltonian with antiperiodic boundary conditions. The same set of
variables $S_k$ are used in both Hamiltonians. To address this problem
we can expand the eigenoperators of one Hamiltonian in terms
of those of the other one. So for example we can write
\bea
D_r&=&\frac{1}{\sqrt{M}}\sum_{k=1}^M e^{-2\pi irk/M}S_k
=\frac{1}{{M}}\sum_{k=1}^M e^{-2\pi irk/M}\sum_{n=0}^{M-1}B_ne^{2\pi ikn/M}
\nonumber\\
&=&\frac{1}{{M}}\sum_{n=0}^{M-1}B_n e^{2\pi i(n-r)/M}\frac{1-e^{2\pi i(n-r)}}{
1-e^{2\pi i(n-r)/M}}=\frac{2}{{M}}\sum_{n=0}^{M-1}B_n \frac{e^{2\pi i(n-r)/M}}{
1-e^{2\pi i(n-r)/M}}\\
{\widetilde D}_r&=&\frac{2}{{M}}\sum_{n=0}^{M-1}{\widetilde B}_n \frac{e^{2\pi i(n-r)/M}}{
1-e^{2\pi i(n-r)/M}}\\
G_r&=&\frac{2}{{M}}\sum_{n=0}^{M-1}\left(B_n\cos\frac{\pi r}{2M}
+i{\widetilde B}_n\sin\frac{\pi r}{2M}\right) \frac{e^{2\pi i(n-r)/M}}{
1-e^{2\pi i(n-r)/M}}\nonumber\\
&=&\frac{2}{{M}}\sum_{n=0}^{M-1}\left(F_n\cos\frac{\pi (n-r)}{2M}
+{\bar F}_n\sin\frac{\pi(n-r)}{2M}\right) \frac{e^{2\pi i(n-r)/M}}{
1-e^{2\pi i(n-r)/M}}\eea 
For nonzero modes ${\bar F}_n=F^\dagger_{M-n}$ so their contribution to
$G_r$ may be written
\bea
\sum_{n=1}^{M-1}\left(F_n C_{rn}+F^\dagger_n S_{rn}\right).
\eea
However the zero mode contribution is
\bea
 \frac{2}{{M}}\left(B_0\cos\frac{\pi r}{2M}
+i{\widetilde B}_0\sin\frac{\pi r}{2M}\right) \frac{e^{-2\pi ir/M}}{
1-e^{-2\pi ir/M}}
\eea 
where $B_0,{\widetilde B}_0$ are anticommuting hermitian operators.
To treat them more like the nonzero modes we define $f_0=(B_0+i{\widetilde B}_0)/2$, so that $\{f_0,f_0^\dagger\}=1$ in terms of which the zero mode contribution
can be written
\bea
&& \frac{2}{{M}}\left(f_0\left(\cos\frac{\pi r}{2M}+\sin\frac{\pi r}{2M}\right)
+f_0^\dagger\left(\cos\frac{\pi r}{2M}-\sin\frac{\pi r}{2M}\right)\right) 
\frac{e^{-2\pi ir/M}}{1-e^{-2\pi ir/M}}\nonumber\\
&&=\sqrt{2}\frac{2}{{M}}\left(f_0\cos\left(\frac{\pi r}{2M}
-\frac{\pi}{4}\right)
+f_0^\dagger\cos\left(\frac{\pi r}{2M}+\frac{\pi}{4}\right)\right) 
\frac{e^{-2\pi ir/M}}{1-e^{-2\pi ir/M}}
\eea 
Then we rescale $F_n=f_n\sqrt{2}$ so that $G_r$ becomes
\bea
G_r&=&\sqrt{2}\sum_{n=0}^{M-1}\left(f_n C_{rn}+f^\dagger_n S_{rn}\right).
\eea
where the matrices $C$ and $S$ are defined by this equation:
\bea
C_{rn}&=&\frac{2}{M}\cos\frac{\pi(n-r)}{2M}\frac{e^{2\pi i(n-r)/M}}{1-
  e^{2\pi i(n-r)/M}},\qquad n\neq0\\
S_{rn}&=&\frac{2}{M}\cos\frac{\pi(n+r)}{2M}\frac{e^{-2\pi i(n+r)/M}}{1-
  e^{-2\pi i(n+r)/M}},\qquad n\neq0\\
C_{r0}&=&\frac{2}{M}\cos\left(\frac{\pi}{4}-\frac{\pi r}{2M}\right)
\frac{e^{-2\pi ir/M}}{1-
  e^{-2\pi ir/M}}\\
S_{r0}&=&\frac{2}{M}\cos\left(\frac{\pi}{4}+\frac{\pi r}{2M}\right)
  \frac{e^{-2\pi ir/M}}{1-e^{-2\pi ir/M}}\eea
  To discuss the continuum limit we need to consider $M\to\infty$ in
  four situations:
  $r,n$ fixed, $r,n^\prime\equiv M-n$ fixed, $r^\prime\equiv M-r,n$ fixed,
  and $r^\prime,n^\prime$ fixed.
\bea
r,n\quad{\rm fixed}:&&\nonumber\\
C_{rn}&\to&\frac{2}{-{2\pi i(n-r)}},\qquad n\neq0\\
S_{rn}&\to&\frac{2}{{2\pi i(n+r)}},\qquad n\neq0\\
C_{r0}&\to&\frac{\sqrt{2}}{{2\pi ir}}\\
S_{r0}&\to&\frac{\sqrt{2}}{{2\pi ir}}
\eea
\bea
r^\prime,n^\prime\quad{\rm fixed}:&&\nonumber\\
C_{rn}&\to&\frac{2}{{2\pi i(n^\prime-r^\prime)}},\qquad n^\prime\neq0\\
S_{rn}&\to&\frac{2}{{2\pi i(n^\prime+r^\prime)}},\qquad n^\prime\neq0\\
C_{r0}&\to&-\frac{\sqrt{2}}{{2\pi ir^\prime}}\\
S_{r0}&\to&\frac{\sqrt{2}}{{2\pi ir^\prime}}
\eea
\bea
r,n^\prime\quad{\rm fixed}:&&\nonumber\\
C_{rn}&\to&\frac{1}{{2 iM}}\to0,\qquad n^\prime\neq0\\
S_{rn}&\to&-\frac{1}{{2 iM}}\to0,\qquad n^\prime\neq0\\
C_{r0}&\to&\frac{\sqrt{2}}{{2\pi ir}}\\
S_{r0}&\to&\frac{\sqrt{2}}{{2\pi ir}}
\eea
\bea
r^\prime,n\quad{\rm fixed}:&&\nonumber\\
C_{rn}&\to&-\frac{1}{{2 iM}}\to0,\qquad n\neq0\\
S_{rn}&\to&-\frac{1}{{2\pi iM}}\to0,\qquad n\neq0\\
C_{r0}&\to&-\frac{\sqrt{2}}{{2\pi ir^\prime}}\\
S_{r0}&\to&\frac{\sqrt{2}}{{2\pi ir^\prime}}
\eea
\subsection{AP Ground state}
Let us first construct the ground state in the AP sector in terms of P
sector states. Then we have to solve $G_r\ket{G}=0$ for all $r$. We make
the ansatz
\bea
\ket{G}&\propto&\exp\left\{\frac{1}{2}f^\dagger_mD_{mn}f^\dagger_n \right\}
\ket{0}\eea
where $f_n\ket{0}=0$ for $n=0,1,2,\ldots$.
Then, if we require $D$ to be antisymmetric, $G_r\ket{G}=0$
implies
\bea
S_{rm}+C_{rn}D_{nm}=0 .
\eea
We attempt to solve for $D$ in the continuum limit.

The equation breaks into 6 equations since $r$ is either near $0$ or near $M$
and $m$ equals 0, is near 0, or is near $M$.
We first write out the equation for
$r$ near 0 and $m=0$:
\bea
S_{r0}+C_{r0}D_{00}+C_{rn}D_{n0}+ C_{rn^\prime}D_{n^\prime 0}=0
    \eea
    where the prime indicates that the index is near $M$ and the zero
    index is explicitly singled out. Putting in the limiting forms and
    cancelling a factor of $2\pi i$ leads to
    \bea
    \frac{\sqrt{2}}{r}&=&\sum_{n=1}^\infty\frac{2}{n-r}D_{n0}
      \eea
      where we used antisymmetry, which implies $D_{00}=0$.
      For $m\neq0$ but near 0, the equation reads
      \bea
S_{rm}+C_{r0}D_{0m}+C_{rn}D_{nm}+ C_{rn^\prime}D_{n^\prime m}=0
\eea
and inserting the limiting forms
 \bea
 \frac{{2}}{r+m}&=&-\frac{\sqrt{2}}{r}D_{0m}
 +\sum_{n=1}^\infty\frac{2}{n-r}D_{nm},\qquad m\neq0.
      \eea
Keeping $r$ near 0 but letting $m$ be near $M$, we write out
      \bea
S_{rm^\prime}+C_{r0}D_{0m^\prime}+C_{rn}D_{nm^\prime}+ C_{rn^\prime}D_{n^\prime m^\prime}=0
\eea
Putting in the limiting forms these equations reduce to
\bea
 0&=&\frac{\sqrt{2}}{r}D_{0m^\prime}
 -\sum_{n=1}^\infty\frac{2}{n-r}D_{nm^\prime}
 \eea
 The remaining 3 equations come from taking $r$ near $M$. The limiting forms of
 the $C$'s all change sign when $r^\prime$ is substituted for $r$:
 \bea
 \frac{\sqrt{2}}{r^\prime}&=&-\sum_{n^\prime=1}^\infty
 \frac{2}{n^\prime-r^\prime}D_{n^\prime 0}\\
  \frac{{2}}{r^\prime+m^\prime}&=&\frac{\sqrt{2}}{r^\prime}D_{0m^\prime}
 +\sum_{n^\prime=1}^\infty\frac{2}{n^\prime-r^\prime}D_{n^\prime m^\prime}\\
 0&=&\frac{\sqrt{2}}{r^\prime}D_{0m}
 -\sum_{n^\prime=1}^\infty\frac{2}{n^\prime-r^\prime}D_{n^\prime m}
  \eea
      We next derive a pair of useful identities following a method
      developed by J. Goldstone \cite{goldstone}.  

      Introduce a meromorphic function $g(z)=\Gamma(z)/\Gamma(z+1/2)$:
      it has poles at  $z=0,-1,-2,\ldots$ and zeroes at $z=-1/2,-3/2,\dots$.
      Near $z=-m$,
      \bea
      g(z)\sim \frac{1}{z+m} \frac{1}{\pi mg(m)},\qquad m=0,1,2,\ldots
      \eea
      Here, for the case $m=0$, we replace $\pi mg(m)\to\sqrt{\pi}$.
      The function $g(z)$  behaves as $Cz^{-1/2}$ as $z\to\infty$,
      Because of the zeroes,
      $g(z)/(z+r)$ has the same poles as $g(z)$ as long as r is a positive
      half odd integer. We therefore can expand
      \bea
      \frac{g(z)}{z+r}&=&\sum_{n=0}^\infty\frac{1}{z+n}\frac{1}{\pi ng(n)}
      \frac{1}{r-n} 
      \label{masterid1}\eea
      where it is important that the left side vanishes as $z\to\infty$,
      Putting $z=m>0$ this becomes
 \bea
 \frac{1}{m+r}&=&\sum_{n=0}^\infty\frac{1}{m+n}\frac{1}{\pi ng(n)g(m)}
      \frac{1}{r-n},
 \qquad m=1,2,\ldots 
      \eea
      which resembles the equations we wish to solve.

      We can get another identity
      by expanding $zg(z)/(z+r)$.
      \bea
      \frac{zg(z)}{z+r}&=&\sum_{n=1}^\infty\frac{-n}{z+n}\frac{1}{\pi ng(n)}
      \frac{1}{r-n}
      \label{masterid2}\\ 
      \frac{1}{m+r}&=&\sum_{n=1}^\infty\frac{-n}{m+n}\frac{1}{\pi ng(n)mg(m)}
      \frac{1}{r-n},
      \quad m=0,1,2,\ldots
      \eea
      In the second equation we put $z=m$. In this case, $m$
      is allowed to be 0, because
      $zg(z)\to1/\sqrt{\pi}$ is finite as $z\to0$.
      The difference of the two
      identities cancels the inhomogeneous term. This is
      useful in solving for the mixed matrix elements
      $D_{n^\prime m}$ and $D_{nm^\prime}$:
      \bea
      0&=&\sum_{n=0}^\infty\frac{1}{\pi ng(n)mg(m)}
      \frac{1}{r-n}=\frac{1}{\sqrt{\pi}mg(m)}\frac{1}{r}
      +\sum_{n=1}^\infty\frac{1}{\pi ng(n)mg(m)}\frac{1}{r-n},
      \qquad m=1,2,\ldots
      \eea
      It is also useful in analyzing ambiguities in the solution of the
      equations. However, in the case of $D_{mn}$, the
      ambiguities in the determination of $D_{mn}$
      are fixed by simply requiring antisymmetry $D^T=-D$.
      
      Comparing to the equations involving these mixed
      matrix elements, we learn that
      \bea
      \frac{D_{nm^\prime}}{D_{0m^\prime}}&=&\frac{1}{\sqrt{2\pi}ng(n)},
      \qquad n>0\\
      \frac{D_{n^\prime m}}{D_{0m}}&=&\frac{1}{\sqrt{2\pi}n^\prime
        g(n^\prime)},      \qquad n^\prime>0
       \eea
       The zero index components aren't determined from these equations
       since the equations are homogeneous.
       They are fixed by the two inhomogeneous
       equations.

      By comparing the identities to the inhomogeneous equations to be solved,
      it is straightforward to infer that
      \bea
      D_{nm}&=&-D_{mn}=-\frac{m-n}{m+n}\frac{1}{2\pi ng(n)mg(m)}.
      \qquad m,n>0\\
      D_{n0}&=&-D_{0n}=\frac{1}{\sqrt{2\pi}ng(n)}\qquad n>0.    
     \eea
     where the antisymmetry of $D$ is assured by taking the sum of the two
     identities for $m>0$, 
     \bea
     \frac{2}{m+r}&=&\frac{1}{\sqrt{\pi}mg(m)}\frac{1}{r}+\sum_{n=1}^\infty\frac{m-n}{m+n}\frac{1}{\pi ng(n)mg(m)}
      \frac{1}{r-n},
      \qquad m=1,2,\ldots
      \eea
      and by using only the second identity for $m=0$,
      \bea
      \frac{1}{r}&=&-\sum_{n=1}^\infty\frac{1}{\sqrt{\pi}ng(n)}
      \frac{1}{r-n}.
      \eea
      The components $D_{n^\prime m^\prime}$ and
      $D_{0n^\prime}$ are the negatives of $D_{nm}$, $D_{0n}$,
      with $n^\prime, m^\prime$ substituted for $n,m$.

      These results are for the continuum limit $M\to\infty$. To assess
      convergence rates we use MATLAB to numerically evaluate the first
      few matrix elements of $D=-C^{-1}S$, which we collect in table
      \ref{numatrix}.
\begin{table}[ht]
\begin{center}
  \begin{tabular}{|c||c|c|c|c|c|
    }
\hline 
    Elements &M=25&M=50&M=200&M=400
    &$M=\infty$\\
\hline
    $D_{10}$&$0.3511-0.0444i$& $0.3529-0.0222i$& $0.3535-0.0056i$&$0.3535-0.0028i$
    &$\sqrt{2}/4$\\
    $D_{20}/D_{10}$&$0.7468-0.0943i$ & $0.7492-0.0471i$& $0.7499-0.0118i$&$0.7500-0.0059i$
    &3/4\\
    $D_{30}/D_{20}$&$0.8320-0.1051i$& $0.8330-0.0524i$& $0.8333-0.0131i$&$0.8333-0.0065i$
    &$5/6$\\
    $D_{40}/D_{30}$&$0.8759-0.1107i$& $0.8752-0.0551i$& $0.8750-0.0137i$&$0.8750-0.0069i$
    &$7/8$\\
    $D_{50}/D_{40}$&$0.9034-0.1141i$& $0.9008-0.0567i$& $0.9001-0.0141i$&$0.9000-0.0071i$
    &$9/10$\\
    $D_{21}$&$0.0294-0.0115i$& $0.0308-0.0059i$& $0.03122-0.00147i$&$0.03124-0.00074i$
    &$1/32$\\
  \hline
\end{tabular} 
\caption{Numerical evaluation of some finite $M$ matrix elements of $D$.
  The imaginary parts tend to zero, in the continuum limit, as $1/M$. The last
column shows the continuum limit $M\to\infty$ obtained in this paper.}
\label{numatrix}
\end{center}
\end{table}

\subsection{AP Excited states}     
To handle excited states we adapt the procedures developed by Sun
\cite{songge} for excited states in
the more complicated three chain transition $P\to P+P$.
The raising operators for AP energy eigenstates are
\bea
g_r^\dagger\equiv \frac{G_r^\dagger}{\sqrt{2}}&=&\sum_{n=0}^{M-1}\left(f_n^\dagger C_{rn}^*+f_n
  S_{rn}^*\right)
\eea
Then we can generate all AP energy eigenstates
from the ground state via the coherent
states
\bea
e^{\sum_r \beta_r g_r^\dagger}\ket{G}&=&
\exp\left\{\frac{1}{2} f_n^\dagger D_{nl}f_l^\dagger+\beta_r K_{rn}
  f_n^\dagger+\frac{1}{2}\beta_r K_{rn}S_{sn}^*\beta_s\right\}\ket{0}
[\det(I+DD^\dagger)]^{-1/4}\nonumber\\
&=&\exp\left\{\frac{1}{2} f^{\dagger T}Df^\dagger+\beta^T Kf^\dagger
+\frac{1}{2}\beta^TKS^{*T}\beta\right\}\ket{0}[\det(I+DD^\dagger)]^{-1/4}\\
K_{rn}&\equiv& C_{rn}^*+S_{rl}^*D_{ln}
\eea
The anticommutation relations of the $G$'s imply the conditions
\bea
 SS^\dagger+CC^\dagger=I,\qquad SC^T+C S^T=0 .
\eea
A short calculation shows that
\bea
CK^T=I
\eea
or in components $C_{rn}K_{ns}=\delta_{rs}$. That is $K^T$ is a right inverse
of $C$. The matrix ${\hat D}\equiv KS^\dagger$ is antisymmetric
as another calculation quickly shows.

The same matrices figure in the inverse transition relating S eigenoperators
to AP eigenoperators:
\bea
f_n&=&\sum_r (g_r {\hat C}_{nr}+g_r^\dagger{\hat S}_{nr})\\
{\hat C}&=&C^\dagger,\qquad {\hat S}=S^T
\eea
The relation between ${\hat C}, {\hat S}$ and $C,S$ follows from direct
comparison of the mode expansions. More generally,
one easily sees that consistency
of the equations $g=Cf+Sf^\dagger$ and $f={\hat C}g+{\hat S}g^\dagger$ leads
to the conditions
\bea
C{\hat C}+S{\hat S}^*&=&I,\qquad C{\hat S}+S{\hat C}^*=0,
\eea
which are solved by ${\hat C}=C^\dagger$ and ${\hat S}=S^T$, although this
  argument doesn't assure uniqueness.

Preservation of anticommutation relations then imposes the further constraints
\bea
C^\dagger C+S^T S^*=I,\qquad S^T C^*+C^\dagger S=0.
\eea
One can then show that $K^T$ is also the left inverse of $C$
(i.e. $K^TC=I$) as follows\footnote{For
  invertible finite dimensional square matrices this is automatic, but these considerations show that we should require this even in the continuum limit when the matrices become infinite dimensional.
}:
\bea
K^TC&=&[C^TK]^T=[(C^\dagger C)^* +C^TS^* D]^T=[(C^\dagger C)^* -S^\dagger
  C D]^T
  \nonumber\\
  &=&[(C^\dagger C)^* +S^\dagger S]^T=I
  \eea
\subsubsection{Continuum limit of $K$}
As before taking the continuum limit
requires breaking the equation into cases: $rs$, $rs^\prime$,$r^\prime s$,
and $r^\prime s^\prime$:
\bea
C_{r0}K_{0s}+C_{rn}K_{ns}+C_{rn^\prime}K_{n^\prime s}&=&\delta_{rs}\\
C_{r^\prime 0}K_{0s}+C_{r^\prime n}K_{ns}+C_{r^\prime n^\prime}K_{n^\prime s}&=&0\\
C_{r0}K_{0s^\prime}+C_{rn}K_{ns^\prime}+C_{rn^\prime}K_{n^\prime s^\prime}&=&0\\
C_{r^\prime0} K_{0s^\prime}+C_{r^\prime n}K_{ns^\prime }
+C_{r^\prime n^\prime}
K_{n^\prime s^\prime }&=&\delta_{r^\prime s^\prime }
\eea
Putting in the limiting forms, these become
\bea
\frac{\sqrt{2}}{r}K_{0s}-\sum_{n=1}^\infty\frac{2}{n-r}K_{ns}&=&2\pi i
  \delta_{rs}\\
  -\frac{\sqrt{2}}{r^\prime}K_{0s}+\sum_{n^\prime=1}^\infty\frac{2}{n^\prime-r^\prime}
  K_{n^\prime s}&=&0\\
  \frac{\sqrt{2}}{r}K_{0s^\prime}-\sum_{n=1}^\infty
  \frac{2}{n-r}K_{ns^\prime}&=&0\\
-\frac{\sqrt{2}}{r^\prime} K_{0s^\prime}
+\sum_{n^\prime=1}^\infty\frac{2}{n^\prime-r^\prime}
K_{n^\prime s^\prime }&=&2\pi i\delta_{r^\prime s^\prime }
\eea
We shall need some more identities to handle the continuum limit
with excited states. If we set
$z=-s$ with $s$ a positive half odd integer in (\ref{masterid1})
or (\ref{masterid2}), the left side goes to zero by virtue
of the zeroes of $g(z)$, unless $s=r$. If $s=r$, the zero in $g(z)$ is
cancelled by the explicit pole:
\bea
\frac{g(z)}{z+r}\to -\delta_{rs}\frac{\pi}{g(r)},
\qquad {\rm as}\quad z\to-s
\eea
So we get
\bea
-\delta_{rs}\frac{\pi}{sg(s)}&=&\sum_{n=0}^\infty\frac{1}{-s+n}
\frac{1}{\pi ng(n)}
      \frac{1}{r-n}
\label{derivedid1}\eea
from (\ref{masterid1}), and
\bea
\delta_{rs}\frac{\pi}{g(s)}&=&\sum_{n=1}^\infty\frac{-n}{-s+n}
\frac{1}{\pi ng(n)}\frac{1}{r-n}
\label{derivedid2}
\eea
from (\ref{masterid2}).
Adding (\ref{derivedid2}) to $s \times$ (\ref{derivedid1}), gives
\bea
0&=&{}-\frac{1}{\sqrt{\pi}}\frac{1}{r}-\sum_{n=1}^\infty
\frac{1}{\pi ng(n)}\frac{1}{r-n}=-\sum_{n=0}^\infty
\frac{1}{\pi ng(n)}\frac{1}{r-n}
\label{derivedid3}
\eea
where in the last form $ng(n)$ is understood to be $1/\sqrt{\pi}$ when $n=0$.
This last identity can be used to find a solution of the middle two equations
relating $K_{n^\prime s}$ to $K_{0s}$ and $K_{ns^\prime}$ to $K_{0s^\prime}$:
\bea
K_{n^\prime s}&=& K_{0s}\frac{1}{\sqrt{2\pi} n^\prime g(n^\prime)}\\
K_{ns^\prime}&=& K_{0 s^\prime}\frac{1}{\sqrt{2\pi} n g(n)}
\eea
We can also add an $s$ dependent factor times the right side of
(\ref{derivedid3}) to the right side of (\ref{derivedid1}) to rearrange
it:
\bea
2\pi i\delta_{rs}&=&\sum_{n=0}^\infty\frac{2isg(s)+\xi_s(s-n)}{s-n}
\frac{1}{\pi ng(n)}\frac{1}{r-n}
      \label{derivedid1a}
      \eea
      Comparing this identity to the first and last of the four equations
      for $K$,
      we can read off
      \bea
      K_{0s}&=&\frac{2ig(s)+\xi_s}{\sqrt{2\pi}}\\
      K_{ns}&=&\frac{1}{2\pi ng(n)}\left[
        \frac{2isg(s)+\xi_s(s-n)}{s-n}\right]\\
      K_{0s^\prime}&=&-\frac{2ig(s^\prime)+\xi_{s^\prime}}{\sqrt{2\pi}}\\
      K_{n^\prime s^\prime}&=&-\frac{1}{2\pi n^\prime g(n^\prime)}\left[
        \frac{2is^\prime g(s^\prime)+\xi_{s^\prime}(s^\prime-n^\prime)}{
          s^\prime-n^\prime}\right]
      \eea
      In these formulas $\xi_s,\xi_{s^\prime}$ are undetermined sets
      of numbers. It will be convenient to replace $\xi_s=-ig(s)(1-\alpha_s)$
      and $\xi_{s^\prime}=-ig(s^\prime)(1+\alpha_{s^\prime})$
      \bea
      K_{0s}&=&\frac{ig(s)(1+\alpha_s)}{\sqrt{2\pi}}\\
      K_{ns}&=&\frac{ig(s)}{2\pi ng(n)}\left[
        \frac{s+n}{s-n}+\alpha_s\right]\\
      K_{0s^\prime}&=&\frac{ig(s^\prime)(-1+\alpha_{s^\prime})}{\sqrt{2\pi}}\\
      K_{n^\prime s^\prime}&=&\frac{ig(s^\prime)}{2\pi n^\prime g(n^\prime)}
      \left[-\frac{s^\prime+n^\prime}{
          s^\prime-n^\prime}+\alpha_{s^\prime}\right]\\
      K_{n^\prime s}&=& 
      \frac{ig(s)}{{2\pi} n^\prime g(n^\prime)}(1+\alpha_s)\\
K_{ns^\prime}&=& \frac{ig(s^\prime)}{{2\pi} n g(n)}
(-1+\alpha_{s^\prime})
\eea
      We have noted that $K^T$ is a right inverse of $C$,
      At the level of string bits, since $C$ is a finite dimensional
      square matrix, if the inverse exists, i.e. if $\det C\neq0$,
      its inverse is unique and is both a left and right inverse. However.
      in the continuum limit $C$ becomes an infinite dimensional matrix and
      these facts need not hold. We seek to determine the $\xi$'s by insisting
      that $K^T$ is also a left inverse of $C$.

      We start by writing out $K^TC=I$ in components (in the continuum limit);
      \bea
      2\pi i\delta_{mn}&=&\sum_r K_{mr}\frac{2}{r-n},\qquad m,n\neq0
      \label{leftk1}\\
2\pi i\delta_{0n}=0&=&\sum_r K_{0r}\frac{2}{r-n},\qquad n\neq0
      \label{leftk10}\\
      2\pi i\delta_{m0}=0&=&\sum_r K_{mr}\frac{\sqrt{2}}{r}-\sum_{r^\prime}
      K_{mr^\prime}\frac{\sqrt{2}}{r^\prime},\qquad m\neq0\label{leftk2}\\
 2\pi i\delta_{00}=2\pi i&=&\sum_r K_{0r}\frac{\sqrt{2}}{r}-\sum_{r^\prime}
 K_{0r^\prime}\frac{\sqrt{2}}{r^\prime},\label{leftk20}\\
      2\pi i\delta_{m^\prime n^\prime}&=&\sum_{r^\prime} K_{m^\prime r^\prime}
      \frac{-2}{r^\prime-n^\prime}\label{leftk6}
 \eea
 \bea
      0&=&\sum_{r^\prime}K_{mr^\prime}\frac{-2}{r^\prime-n^\prime},
      \qquad m\neq0
      \label{leftk3}\\
           0&=&\sum_{r^\prime}K_{0r^\prime}\frac{-2}{r^\prime-n^\prime}
      \label{leftk30}\\
      0&=&\sum_{r}K_{m^\prime r}\frac{2}{r-n},\qquad n\neq0
      \label{leftk4}\\
0&=&\sum_r K_{m^\prime r}\frac{\sqrt{2}}{r}-\sum_{r^\prime}
      K_{m^\prime r^\prime}\frac{\sqrt{2}}{r^\prime}\label{leftk5}
      \eea
      where we have separated the components with $m$=0 or $n=0$ or both.
      Recall that when an index is primed it is always nonzero.
      
      In order to solve these equations by Goldstone's method,
      we employ $1/g(z)$ instead of $g(z)$. The poles of $1/g$
are at $z=-r$:
\bea
\frac{1}{g(z)}\sim-\frac{rg(r)}{\pi(z+r)},\qquad r=1/2,3/2,\cdots
\eea
and we have the expansion
\bea
\frac{1}{g(z)}\frac{1}{z+n}&=&\sum_{r=1/2}^\infty\frac{-rg(r)}{\pi(z+r)}
\frac{1}{n-r},\qquad n=0,1,2,\cdots
\label{leftzexp1}
\eea
Setting $z=-m$ the left side vanishes for $m\neq n$. For $m=n$,
$g(z)(z+n)\to 1/[\pi ng(n)]$, so we get
\bea
\delta_{mn}{\pi mg(m)}&=&\sum_{r=1/2}^\infty\frac{-rg(r)}{\pi(-m+r)}
\frac{1}{n-r},\qquad m,n=0,1,2,\cdots
\label{leftid1}
\eea
Using $1/[zg(z)]$ instead gives the expansion
\bea
\frac{1}{zg(z)}\frac{1}{z+n}&=&\sum_{r=1/2}^\infty\frac{g(r)}{\pi(z+r)}
\frac{1}{n-r},\qquad n=1,2,\cdots
\label{leftzexp2}
\eea
$n=0$ is excluded because in that case the pole at $z=0$ is not cancelled.
Letting $z\to-m\neq0$, the left side goes to $-\delta_{mn}\pi g(m)$, producing
the identity
\bea
-\delta_{mn}\pi g(m)&=&\sum_{r=1/2}^\infty\frac{g(r)}{\pi(r-m)}
\frac{1}{n-r},\qquad m,n=1,2,\cdots
\label{leftid2}\eea
  Again it is convenient to choose linear combinations of (\ref{leftid1}) and
  (\ref{leftid2}). For both $m,n$ nonzero we can add $m$ times (\ref{leftid2})
    to (\ref{leftid1}) so that the left sides cancel
    \bea
    0&=&-\sum_{r=1/2}^\infty\frac{g(r)}{\pi mg(m)}\frac{1}{n-r},\qquad m,n=1,2,\cdots
    \label{homog1}
    \eea
    Then one can add an $m$ dependent multiple of this last equation to
    (\ref{leftid1}) to put the latter in the form
\bea
2\pi i\delta_{mn}&=&\sum_{r=1/2}^\infty\frac{2ig(r)}{\pi g(m)}
\left[-\frac{r}{m}\frac{1}{r-m}+\frac{\eta_m}{m}\right]\frac{1}{n-r},
\qquad m,n=1,2,\cdots
\label{leftid1a}
\eea
Comparing this last equation to (\ref{leftk1}) and (\ref{leftk6})
allows us to read off the nonzero unmixed index components of $K^T$
\bea
K_{mr}&=&\frac{2ig(r)}{2\pi mg(m)}
\left[\frac{r}{r-m}-{\eta_m}\right],\qquad m\neq0\\
K_{m^\prime r^\prime}&=&\frac{2ig(r^\prime)}{2\pi m^\prime g(m^\prime)}
\left[-\frac{r^\prime}{r^\prime-m^\prime}
  -{\eta_{m^\prime}}\right],\qquad m^\prime\neq0
\eea
The mixed components $K_{mr^\prime}$ and $K_{m^\prime r}$ satisfy
(\ref{leftk3}) or (\ref{leftk4}). Referring to (\ref{homog1}), we see that
these equations can be solved as $m$ or $m^\prime$ dependent factors times
$g(r^\prime)$ or $g(r)$. Information about the $m$ or $m^\prime$ dependence
is contained in (\ref{leftk2}) or (\ref{leftk5}). We compute
\bea
\sum_r K_{mr}\frac{1}{r}&=&\sum_r \frac{ig(r)}{\pi mg(m)}
\left[\frac{1}{r-m}-\frac{\eta_m}{r}\right]
=-\frac{i\eta_m}{\pi mg(m)}\sum_r\frac{g(r)}{r}.
\eea
Inserting this into (\ref{leftk2}) gives
\bea
K_{mr^\prime}&=&-\frac{i\eta_m g(r^\prime)}{\pi mg(m)}
\eea
Similarly (\ref{leftk5}) can be used to determine
\bea
K_{m^\prime r}&=&-\frac{i\eta_{m^\prime} g(r)}{\pi m^\prime g(m^\prime)}
\eea
It is again convenient to replace $\eta_m=(1+\beta_m)/2$ and
$\eta_{m^\prime}=(-1+\beta_{m^\prime})/2$.
\bea
K_{mr}&=&\frac{ig(r)}{2\pi mg(m)}
\left[\frac{r+m}{r-m}-{\beta_m}\right],\qquad m\neq0\\
K_{m^\prime r^\prime}&=&\frac{2ig(r^\prime)}{2\pi m^\prime g(m^\prime)}
\left[-\frac{r^\prime+m^\prime}{r^\prime-m^\prime}
  -{\beta_{m^\prime}}\right],\qquad m^\prime\neq0\\
K_{mr^\prime}&=&-\frac{i(1+\beta_m) g(r^\prime)}{2\pi mg(m)}\\
K_{m^\prime r}&=&
-\frac{i(-1+\beta_{m^\prime}) g(r)}{2\pi m^\prime g(m^\prime)}
\eea
We finally consider $K_{0r}$ and $K_{0r^\prime}$, which satisfy
(\ref{leftk10}) and (\ref{leftk30}) respectively. These equations are solved by
$K_{0r}=C_0g(r)$ and $K_{0r^\prime}=C_0^\prime g(r^\prime)$.
Then the final equation  (\ref{leftk20}) reads
\bea
2\pi i&=&\sqrt{2}(C_0-C_0^\prime)\sum_r\frac{g(r)}{r}.
\eea
The $m=n=0$ case of (\ref{leftid1} reads
\bea
\sqrt{\pi}=\frac{1}{\pi}\sum_r\frac{g(r)}{r}
\eea
so that
\bea
C_0-C_0^\prime&=&i\sqrt{\frac{2}{\pi}}
\eea
We write the solutions as $C_0=i(1-\beta_0)/\sqrt{2\pi}$ and
$C_0^\prime=i(-1-\beta_0)/\sqrt{2\pi}$, completing the solution of the
left inverse equations:
\bea
K_{0r}&=&\frac{ig(r)(1-\beta_0)}{\sqrt{2\pi}}\\
K_{0r^\prime}&=&\frac{ig(r^\prime)(-1-\beta_0)}{\sqrt{2\pi}}
\eea
Equating these solutions to the corresponding
results of the right inverse calculation, leads to the conclusion that
all $\beta$'s have the common value $\beta$, all the $\alpha$'s have the same value $\alpha$, and $\beta=-\alpha$.

By insisting that $K^T$ be both a left and right inverse of $C$ we have
reduced the ambiguities in the solution to the single parameter $\alpha$,
which is not
determined by the the full set of left and right inverse equations.
However. since interchanging primed and unprimed indices reverses
the sign of the
elements of $C$, we can require that this applies also for the elements of $K$.
This then would determine $\alpha=0$.
\bea
K_{mr}&=&\frac{ig(r)}{2\pi mg(m)}
\left[\frac{r+m}{r-m}\right],\qquad m\neq0\nonumber\\
K_{m^\prime r^\prime}&=&-\frac{ig(r^\prime)}{2\pi m^\prime g(m^\prime)}
\left[\frac{r^\prime+m^\prime}{r^\prime-m^\prime}\right]\nonumber\\
K_{0r}&=&\frac{ig(r)}{\sqrt{2\pi}}\nonumber\\
K_{0r^\prime}&=&-\frac{ig(r^\prime)}{\sqrt{2\pi}}\nonumber\\
K_{mr^\prime}&=&-i\frac{g(r^\prime)}{2\pi mg(m)},\qquad m\neq0 \nonumber\\
K_{m^\prime r}&=&i\frac{g(r)}{2\pi m^\prime g(m^\prime)}
\eea
\subsubsection{Continuum limit of ${\hat D}$}
From its definition ${\hat D}\equiv KS^\dagger$ we can quickly show
that ${\hat D}$ is antisymmetric:
\bea
{\hat D}^T&=& S^* K^t=S^*(C^\dagger+D^TS^\dagger)
S^\dagger
=-(C^*+S^8D)S^\dagger=-KS^\dagger=-{\hat D}
\eea
Furthermore we can obtain an equation for ${\hat D}$, similar to the one
satisfied by $D$: 
\bea
{\hat D}C&=&KS^\dagger C=-KC^T S^*=-(CK^T)^TS^*=-S^*
\eea
Then we can search for solutions in the continuum limit.
 \bea
 2\pi i(-S^*_{rn})\to\frac{2}{r+n}
 &=&\sum_s {\hat D}_{rs}\frac{2}{s-n},\qquad n\neq0
      \label{leftdhat1}\\
      \frac{\sqrt{2}}{r}&=&\sum_s {\hat D}_{rs}\frac{\sqrt{2}}{s}
      -\sum_{s^\prime}
      {\hat D}_{rs^\prime}\frac{\sqrt{2}}{s^\prime},
      \label{leftdhat2}\\
      0&=&\sum_{s^\prime}{\hat D}_{rs^\prime}\frac{-2}{s^\prime-n^\prime},
      \label{leftdhat3}\\
      0&=&\sum_{s}{\hat D}_{r^\prime s}\frac{2}{s-n},\qquad n\neq0
      \label{leftdhat4}\\
      \frac{\sqrt{2}}{r^\prime}
        &=&\sum_s {\hat D}_{r^\prime s}\frac{\sqrt{2}}{s}-\sum_{s^\prime}
{\hat D}_{r^\prime s^\prime}\frac{\sqrt{2}}{s^\prime}
\label{leftdhat5}\\
      \frac{{2}}{r^\prime+n^\prime}&=&\sum_{s^\prime}
      {\hat D}_{r^\prime s^\prime}
      \frac{-2}{s^\prime-n^\prime}\label{leftdhat6}
      \eea
      In (\ref{leftzexp1}) change the summation index to $s$ and then
      set $z=r$ to get
      \bea
      \frac{1}{r+n}&=&\sum_{s=1/2}^\infty
      \frac{-sg(s)g(r)}{\pi(r+s)}
\frac{1}{n-s},\qquad n=0,1,2,\cdots
\label{leftdid1}
\eea
Doing the same in (\ref{leftzexp1}) gives a second identity
\bea
\frac{1}{r+n}&=&\sum_{s=1/2}^\infty\frac{rg(r)g(s)}{\pi(r+s)}
\frac{1}{n-s},\qquad n=1,2,\cdots
\label{leftdid2}
\eea
For $n>0$ adding (\ref{leftdid1}) and (\ref{leftdid2}) shows that
\bea
{\hat D}_{rs}&=&\frac{(s-r)g(s)g(r)}{2\pi(r+s)}
\eea
satisfies (\ref{leftdhat1}) while
\bea
{\hat D}_{r^\prime s^\prime}&=&-\frac{(s^\prime-r^\prime)g(s^\prime)
  g(r^\prime)}{2\pi(r^\prime+s^\prime)}
\eea
satisfies (\ref{leftdhat6}).

Also for $n>0$ taking the difference (\ref{leftdid1})
minus (\ref{leftdid2}) gives
\bea
0&=&\sum_{s=1/2}^\infty
      \frac{g(s)g(r)}{\pi}\frac{1}{s-n},
      \eea
      which shows that
      \bea
      {\hat D}_{rs^\prime}&=& \delta_r\frac{g(s^\prime)g(r)}{\pi}\\
      {\hat D}_{r^\prime s}&=& \delta_{r^\prime}\frac{g(s)g(r^\prime)}{\pi}
      \eea
      satisfy (\ref{leftdhat3}) (\ref{leftdhat4}) respectively.

      Finally we consider (\ref{leftdid1}) for $n=0$:
      \bea
      \frac{1}{r}&=&\sum_{s=1/2}^\infty
      \frac{g(s)g(r)}{\pi(r+s)},
      \eea
      which we can use to evaluate
      \bea
      \sum_s {\hat D}_{rs}\frac{\sqrt{2}}{s}&=&\sum_s
      \frac{(s-r)g(s)g(r)}{2\pi(r+s)}\frac{\sqrt{2}}{s}
      =\sum_s\frac{\sqrt{2}g(s)g(r)}{\pi(r+s)}-\sum_s
      \frac{\sqrt{2}g(s)g(r)}{2\pi s}\nonumber\\
      &=&\frac{\sqrt{2}}{r}-g(r)\sum_s\frac{\sqrt{2}g(s)}{2\pi s}
      \eea
      Then (\ref{leftdhat2}) reads
      \bea
      0&=&-g(r)\sum_s\frac{\sqrt{2}g(s)}{2\pi s}-\delta_rg(r)
      \sum_{s^\prime}\frac{\sqrt{2}g(s^\prime)}{\pi s^\prime}
    \eea
    determining $\delta_r=-1/2$. Similarly applying the same analysis to
    (\ref{leftdhat5}) determines $\delta_{r^\prime}=+1/2$. Thus
\bea
      {\hat D}_{rs^\prime}&=& -\frac{g(s^\prime)g(r)}{2\pi}\\
      {\hat D}_{r^\prime s}&=& \frac{g(s)g(r^\prime)}{2\pi}
      \eea
      This completes the determination of all the matrix elements of
      ${\hat D}$ in the continuum limit. The ambiguities are resolved
      because we imposed antisymmetry ${\hat D}^T=-{\hat D}$.

\section{Vertex for 3 antiperiodic strings.}      
We next consider the transition of two smaller chains into one larger chain. 
If all three strings are antiperiodic, it is easy enough to evaluate the overlap matrices
with $M$ finite.
\bea
C_{sr1}&=&-\frac{1}{\sqrt{MK}}
\frac{1+e^{-2\pi isK/M}}{1-e^{2i\pi (s/M-r/K)}}\cos\left(\frac{s\pi}{2M}
  -\frac{r\pi}{2K}\right)\nonumber\\
C_{sr2}&=&-\frac{1}{\sqrt{ML}}
\frac{-1+e^{-2\pi isK/M}}{1-e^{2i\pi (s/M-r/L)}}\cos\left(\frac{s\pi}{2M}
-\frac{r\pi}{2L}\right)\\
S_{sr1}&=&-\frac{1}{\sqrt{MK}}
\frac{1+e^{-2\pi isK/M}}{1-e^{2i\pi (s/M+r/K)}}\cos\left(\frac{s\pi}{2M}
  +\frac{r\pi}{2K}\right)\nonumber\\
S_{sr2}&=&-\frac{1}{\sqrt{ML}}
\frac{-1+e^{-2\pi isK/M}}{1-e^{2i\pi (s/M+r/L)}}\cos\left(\frac{s\pi}{2M}
  +\frac{r\pi}{2L}\right).
\eea
The continuum limit of these matrices is $K,L\to\infty$ with $K/M=x$ with
$0<x<1$ fixed and $s$ or $M-s$, $r1$ or $K-r1$, and $r2$ or $L-r2$
finite. Unlike the cases where the number of antiperiodic strings is even,
when the $C$'s and $S$'s can be reduced to reciprocals of linear combinations
of integers,
there is no common overall factor that can be removed from $C,S$ when the
number of antiperiodic strings is odd and the dependence on exponentials of the
indices cannot be factored off.
We have not been able to modify Goldstone's method
to handle this more complicated situation.

However if the number of bits $M$ stays finite there is nothing to prevent
a numerical solution of the equations for $D$, $K$, and ${\hat D}$. To
illustrate this (and indicate the accuracy of the numerics), we used MATLAB to
calculate the first few elements of $D$, with increasing $M$ at fixed $x=K/M$, in Tables \ref{realD}
and \ref{imD}.
\begin{table}[ht]
\begin{center}
  \begin{tabular}{|c||c|c|c|c|c|c|
    }
    \hline
    M&   Re $100 D_{\frac{3}{2}\frac{1}{2}}$   & Re  $100 D_{\frac{5}{2}
                                                 \frac{1}{2}}$
    & Re $100 D_{\frac{5}{2}\frac{3}{2}}$      &   Re $100 D_{\frac{7}{2}
                                                 \frac{1}{2}}$
    &   Re $100 D_{\frac{7}{2}\frac{3}{2}}$     &  Re $100 D_{\frac{7}{2}
                                                  \frac{5}{2}}$ \\
\hline
14& 2.2447648& 1.3055708& 0&  0& -.51953822& -.36957678\\
28& 2.8152666& 2.5958446&  .54056077&  2.0870933&  .58406856&  .13751509\\
56& 2.9583407& 2.9233054&  .67876263&  2.6263502&  .87204834&  .27324926\\
112& 2.9941314& 3.0054722&  .71347885&  2.7623182&  .94472719&  .30761115\\
224& 3.0030800& 3.0260318&  .72216757&  2.7963820&  .96293806& .31622643\\
448& 3.0053171& 3.0311728&  .72434031&  2.8049023&  .96749326&  .31838175\\
896& 3.0058764& 3.0324580&  .72488353&  2.8070326&  .96863220&  .31892066\\
1792& 3.0060162& 3.0327794&  .72501933&  2.8075652&  .96891695&  .31905540\\
\hline
\end{tabular}
\caption{The real parts of the first few elements of $100D$, with both indices on
  short string 1, for the
  vertex of 3 antiperiodic
  strings for the case
$x=K/M=4/7$ and increasing values of $M$.}
\label{realD}
\end{center}
\end{table}
\begin{table}[ht]
\begin{center}
  \begin{tabular}{|c||c|c|c|c|c|c|
    }
    \hline
 M&   Im-$100 D_{\frac{3}{2}\frac{1}{2}}$   & Im-$100 D_{\frac{5}{2}\frac{1}{2}}$
    & Im-$100 D_{\frac{5}{2}\frac{3}{2}}$      &   Im-$100 D_{\frac{7}{2}\frac{1}{2}}$
    & Im-$100 D_{\frac{7}{2}\frac{3}{2}}$     &  Im-$100 D_{\frac{7}{2}\frac{5}{2}}$ \\
\hline
14&  2.2447648&3.1519267&    .89896526&3.4549332&1.2542762&.36957678\\
28& 1.1661216&1.7344879&   .54056077&2.0870933&.87412037&.33199081\\
56& .58845055&.88677499&   .28115269&1.0878699&.46611955&.18257932\\
112& ,29489620&.44581968&   .14191977&.54945926&.23664184&.09331282\\
224&.14753186&.22321352&  .07112730&.27541959&.11876708&.04690776\\
448&.07377639&.11164470&  .03558456&.13779601&.05943926&.02348525\\
896&.03688950&.05582709&  .01779489&.06890878&.02972659&.01174654\\
1792&.01844492&.02791414&  .00889777&.03445574&.01486417&.00587376\\
\hline
\end{tabular}
\caption{The imaginary parts of the first few elements of $100\times D$,
  with both indices on short string 1,
  for the vertex of 3 antiperiodic strings for the case
$x=K/M=4/7$. Notice that the imaginary parts tend to 0 as $1/M$.}
\label{imD}
\end{center}
\end{table}
For an analytic treatment of the continuum limit, in the absence of an
adaptation of Goldstone's method
to this situation, we have the method described in the introduction
and illustrated in Fig. \ref{apappap}, and a less direct approach
exploiting bosonization explained in the following subsection
\subsection{Bosonization}
Another tool to evaluate vertex overlaps, in the continuum limit,
is to double the number of
real fermion fields and describe the resulting system in the language
of bosonization. Since the boson fields are free, one can then borrow
Mandelstam's results for the overlap for bosonic coordinates
and then express them in terms of
the fermion fields using the explicit formulas of bosonization. It is
convenient to combine the pair of real fermion fields into a single
complex fermion field using $\psi=(S^1+iS^2)/\sqrt{2}$, or in terms of modes
$b=(b^1+ib^2)/\sqrt{2}$ , $d=(b^1-ib^2)/\sqrt{2}$, where $b$ destroys 1 unit of positive charge and $d$ destroys one unit of negative charge. Then for the
three vertex the ground state of the long string has the structure
\bea
\ket{G}&\propto&\exp\left\{b^\dagger_rD_{rs}d^\dagger_s \right\}
\ket{0}\eea
where the indices $r,s$ run over the (half integer) modes of the two
smaller strings.The structure in bosonized language is \cite{mandelstamlc}
\bea
\ket{G}&\propto&\exp\left\{\frac{1}{2}\sum_{m,n>0}a_{-m}N_{mn}a_{-n}
+P\sum_{m>0}N_ma_{-m}-\tau_0\frac{P^2}{2\alpha_1\alpha_2\alpha_3}\right\}
\nonumber\\
&&\exp\left\{\frac{1}{2}\sum_{m,n>0}{\tilde a}_{-m}N_{mn}{\tilde a}_{-n}
  +{\tilde P}\sum_{m>0}N_m{\tilde a}_{-m}
  -\tau_0\frac{{\tilde P}^2}{2\alpha_1\alpha_2\alpha_3}\right\}\ket{0}
\eea
where $m,n$ run over the bosonic modes of the two shorter strings.
$\tau_0$ depends only on the $\alpha$'s:
\bea
\tau_0&=&\alpha_1\ln\left|\frac{\alpha_1}{\alpha_3}\right|
  +\alpha_2\ln\left|\frac{\alpha_2}{\alpha_3}\right|=|\alpha_3|[x\ln x
  +(1-x)\ln(1-x)].
\eea
The symbol $P$
contains the zero mode information
\bea
P=\alpha_1 a_0^2-\alpha_2a_0^1,\qquad
{\tilde P}=\alpha_1 {\tilde a}_0^2-\alpha_2{\tilde a}_0^1 
\eea
where $\alpha_i=2P^+_i>0$ for $i=1,2$ where $i$ labels the two short strings.
The long string is labeled by $i=3$ and $\alpha_3=-\alpha_1-\alpha_2$ so
$P^+$ conservation means $\sum_{r=1}^3\alpha_r=0$. While an uncompactified
string coordinate requires $a_0={\tilde a}_0$, the bosonic coordinate,
describing a spin system, is compactified and this equality need not hold. 
We shall first assume that $P={\tilde P}=0$.
This corresponds in fermion language to restricting the
states to be neutral in both the charge and the chiral charge. In the
boson language charge corresponds to Kaluza-Klein (KK) momenta and chiral
charge to winding number. The dependence on them will be determined
later. The $N_{mn}, N_m$ are all known from Mandelstam's
work on interacting strings \cite{mandelstamlc}. we quote them using
Goldstone's $g(z)$ function
\bea
g(z)&=&\frac{\Gamma(1+xz)\Gamma(1+(1-x)z)e^{z\xi}}{z\Gamma(1+z)\sqrt{x(1-x)}}
\\
\xi&\equiv&-\frac{\tau_0}{\alpha_1+\alpha_2};
\eea
Making the appropriate modifications\footnote{
  Mandelstam's original $N$'s were obtained for open strings. For closed
  strings, the phases of the $N$'s depend on the $\sigma$'s chosen for
  the interaction points. The following $N$'s correspond to the $\sigma$ for
  string 1 identified with the interval $0<\sigma<P_1^+$ on the long string,
  with the $\sigma$ for string 2 identified with the interval
  $P_1^+<\sigma<P^+$ on the  long string. The open string $N$'s are
  appropriate for the closed string parameterization chosen in
  \cite{greenschwarzbrink}
}
to apply to our choice of
parameterization, The $N$'s occurring in our bosonic formula are
\bea
N^1_m&=&-\frac{1}{\alpha_1}\frac{x^{3/2}}{m^2g(m/x)\sqrt{1-x}}\nonumber\\
N^2_m&=&\frac{1}{\alpha_2}\frac{(1-x)^{3/2}}{m^2g(m/(1-x))\sqrt{x}}\nonumber\\
N^{11}_{mn}&=&\frac{x}{(m+n)mg(m/x)ng(n/x)}\\
N^{22}_{mn}&=&\frac{1-x}{(m+n)mg(m/(1-x))ng(n/(1-x))}\\
N^{12}_{mn}&=&-\frac{x(1-x)}{(m(1-x)+nx)mg(m/x)ng(n/(1-x))}\\
\eea
Mandelstam's $N$'s have extra sign factors $(-)^m$ for $N^1_m$ and
$N^{12}_{mn}$  and $(-)^{m+n}$ for $N^{11}_{mn}$.

The explicit bosonization formula is
\bea
a_{-m}=a_m^\dagger&=&\sum_{r=1/2}^\infty \left[b^\dagger_{r+m} b_r-
  d^\dagger_{r+m} d_r\right] +\sum_{r=1/2}^{m-1/2}b^\dagger_{m-r}d^\dagger_r
\eea
with a corresponding formula relating ${\tilde a}_m$ to ${\tilde b}_r,
{\tilde d}_r$.
One finds, with due care with operator ordering,
\bea
[a_m,a_n]&=& m\delta_{m+n,0},\qquad [a_m,{\tilde a}_n]=0,\qquad
[{\tilde a}_m,{\tilde a}_n]= m\delta_{m+n,0}
\eea
We see that the zero mode
\bea
a_0=\sum_{r=1/2}^\infty \left[b^\dagger_{r} b_r-d^\dagger_{r} d_r\right]
\eea
is just the charge carried by the $b,d$ operators and ${\tilde a}_0$ is
the charge carried by the ${\tilde b},{\tilde d}$ operators, and
there are certainly states for which these charges are different. The KK
momentum is $a_0+{\tilde a}_0$ and the winding number is $a_0-{\tilde a}_0$.

One can now consider the vertex for the particular states
\bea
a^\dagger_m a^\dagger_n\ket{0}&=&\sum_{s=1/2}^{n-1/2}
\left(b^\dagger_{m+n-s}d^\dagger_s-b^\dagger_sd^\dagger_{m+n-s}\right)\ket{0}
+\left(\sum_{s=1/2}^{n-1/2}b^\dagger_{n-s}d^\dagger_s\right)
\left(\sum_{r=1/2}^{m-1/2}b^\dagger_{m-r}d^\dagger_r\right)\ket{0}\eea
Plugging the left side into Mandelstam's formula yields $mnN_{mn}$.
On the other hand plugging the right side into the assumed structure
for the fermion language yields
\bea
mnN_{mn}&=&\sum_{s=1/2}^{n-1/2}
\left(D_{m+n-s,s}-D_{s,m+n-s}\right)
+2\sum_{s=1/2}^{n-1/2}\sum_{r=1/2}^{m-1/2}\left(D_{n-s,s}D_{m-r,r}-
D_{n-s,r}D_{m-r,s}\right)
\eea
This formula determines recursively the $D$'s in terms of the $N$'s.
In particular, for $x=4/7$ (the case tabulated in the tables)
\bea
D^{11}_{3/2,1/2}&=&\frac{1}{2}N^{11}_{11}
=\frac{1}{4}\left[\frac{3}{7}\right]^{5/2}\approx 0.0300606\\
D^{11}_{5/2,1/2}&=&N^{11}_{21}=\frac{10}{21}\left[\frac{3}{7}\right]^{13/4}
\approx0.0303289\eea
which are very close to the last entries in Table \ref{realD}, for $M=1792$.
Here the superscripts
refer to which of the two shorter strings the mode numbers $r,s$ belong to.
The table is compiled
assuming they both refer to string 1.

The quadratic terms of the recursion
formula do not contribute in the determination of $D^{11}_{3/2,1/2}$
and $D^{11}_{5/2,1/2}$. That changes for larger values of the indices.
Indeed, by studying the state $a_{-3}a_{-1}\ket{0}$ one learns that
  \bea
  D_{7/2,1/2}&=&\frac{3}{2}N^{11}_{31}-\frac{1}{2}(D_{3/2,1/2})^2
  \nonumber\\
  &=&\frac{3\cdot13\cdot17}{4\cdot49}\left(\frac{3}{7}\right)^{3/2}D_{3/2,1/2}
  -\frac{1}{2}(D_{3/2,1/2})^2\approx0.0280774
  \eea
again very close to the tabulated result for $M=1792$.

So far we have considered only states with $P={\tilde P}=0$.
To check on the zero mode structure of the bosonic formalism we need to
study short string states
with nonzero $a_0$ and/or nonzero ${\tilde a}_0$. For example the
states
\bea
b_{1/2}^\dagger b_{3/2}^\dagger b_{7/2}^\dagger\cdots b_{n-1/2}^\dagger\ket{0} 
\eea
have $a_0=n>0$ and the corresponding states with $b\to d$ have $a_0=-n<0$,
with similar constructions for the tilde operators. Since there are no gaps
in the string of operators, it is easy to show that all these states
are annihilated by $a_n,{\tilde a}_n$ with $n>0$. This means that if the
short string states are selected from these, the only state dependence of the
bosonic vertex is given by the zero mode factors
\bea
\exp\left\{-\tau_0\frac{P^2}{2\alpha_1\alpha_2\alpha_3}
-\tau_0\frac{{\tilde P}^2}{2\alpha_1\alpha_2\alpha_3}\right\}
\eea
Since the ground state of the long string has $a_0={\tilde a}_0=0$
the two short strings must have opposite values of these charges.

\begin{table}[ht]
\begin{center}
  \begin{tabular}{|c||c|c|
    }
    \hline
 M&   ${\rm Re} D^{12}_{{1}/{2},{1}/{2}}$&${\rm Re} D^{12}_{{3}/{2},{1}/{2}}$ \\
\hline
14&  -0.11078489&-0,10214029\\
28&  -0.056457379&-0.054671329\\
56&  -0.028362255&-0.027791394\\
112& -0.014197838&-0.013952835\\
224& -0.0071010083&-0.0069835650\\
448& -0.0035507653&-0.0034926761\\
896& -0.0017754153&-0.0017464497\\
1792&-0.00088771173&-0.00087323883\\
    \hline
\end{tabular}
\caption{Two elements of the matrix ${\rm Re} D$
  with the first index $=1/2$ or$=3/2$ on short string 1 and the second $=1/2$
  on short string 2,
  for the vertex of 3 antiperiodic strings for the case $x=K/M=4/7$.
Notice that the real parts vanish as $1/M$.}
\label{reD12}
\end{center}
\end{table}
\begin{table}[ht]
\begin{center}
  \begin{tabular}{|c||c|c|
    }
    \hline
 M&   ${\rm Im} D^{12}_{{1}/{2},{1}/{2}}$&${\rm Im} D^{12}_{{3}/{2},{1}/{2}}$ \\
\hline
14&   0.22464955& 0.089574631\\
28&   0.24213249& 0.12066219\\
56&   0.24654097& 0.12867619\\
112&  0.24764546& 0.13069489\\
224&  0.24792173& 0.13120051\\
448&  0.24799081& 0.13132697\\
896&  0.24800808& 0.13135859\\
1792& 0.24801240& 0.13136650\\
    \hline
\end{tabular}
\caption{Two elements of the matrix ${\rm Im}D$
  with the first index $=1/2$ or $=3/2$ on short string 1 and the second $=1/2$
  on short string 2,
  for the vertex of 3 antiperiodic strings for the case $x=K/M=4/7$.
  Here the imaginary parts converge to a nonzero limit
  for large $M$.}
\label{imD12}
\end{center}
\end{table}
Let's take, as a simple example, the two short string state
$\ket{0;1,-1}=b^{1\dagger}_{1/2}
d^{2\dagger}_{1/2}\ket{0}e^{i\eta}$,
which has $a_0^1=-a_0^2=1$ and ${\tilde a}_0=0$. We included a phase factor,
which is not fixed by specifying only the charges of the two strings.
For this state ${\tilde P}=0$ and $P=-(\alpha_1+\alpha_2)=\alpha_3$. Since
this is the lowest energy two string state with these charges,
$a_n\ket{0;1,-1}=0$ for all $n>0$.
Then
\bea
\exp\left\{-\tau_0\frac{P^2}{\alpha_1\alpha_2\alpha_3}\right\}
  &=&x^{1/(1-x)/2}(1-x)^{1/x/2}\to (4/7)^{7/6}(3/7)^{7/8}
\approx0.24801384
\eea
for the value $x=4/7$ which was used in the tabulations.
Tables~\ref{reD12}, \ref{imD12} show the values for the matrix elements
$D^{12}_{1/2,1/2}$ and $D^{12}_{3/2,1/2}$
for increasing values of $M$ in the string bit model. The bosonic
calculation is quite
close to ${\rm Im} D^{12}_{1/2,1/2}$ in the last entry of Table \ref{imD12}.
The corresponding real part shown in Table \ref{reD12} 
is also very small. In other words there is excellent
agreement if the phase of the state is chosen to be $e^{i\eta}=e^{-i\pi/2}=-i$.
Moreover, from the
equations satisfied by $D$ and the explicit forms for $C$ and $S$, one
can argue that, in the continuum limit,
the elements of $D^{11}$ and $D^{22}$ are all real, while the
elements of $D^{12}$ are all imaginary. Thus we can anticipate that
we should identify the states
\bea
\ket{0;n,-n}&=&\pm b_{1/2}^{1\dagger} d_{1/2}^{2\dagger}
b_{3/2}^{1\dagger} d_{3/2}^{2\dagger}\cdots
b_{n-1/2}^{1\dagger} d_{n-1/2}^{2\dagger}\ket{0}e^{-in\pi/2}
\eea
where the overall sign can be settled by comparing the analytic bosonic
calculation to the
numerical fermionic calculation for large $M$.
Since the bosonization formulas don't fix the
relative phase of states of different strings, this is satisfactory.

We have also tabulated the element $D^{12}_{3/2,1/2}$ which is determined by
the state $b^{1\dagger}_{3/2}d^{2\dagger}_{1/2}\ket{0}$.
To use bosonization to calculate the
$M\to\infty$ limit, we first note that
\bea
a^1_{-1}\ket{0;1,-1}&=&-ia^1_{-1}b^{1\dagger}_{1/2}d^{2\dagger}_{1/2}\ket{0}
\nonumber\\
&=&-ib^{1\dagger}_{3/2}d^{2\dagger}_{1/2}\ket{0}\eea
The contribution of the last line to the long string ground state is
just $-iD^{12}_{3/2,1/2}$. In bosonized language the left side of the
first line shows that it is $PN^1_1(-iD^{12}_{1/2,1/2})$. Thus we conclude that
\bea
D^{12}_{3/2,1/2}&=&PN^1_1D^{12}_{1/2,1/2}.
\eea
For the value of $x=4/7$ chosen for the tables, $P=\alpha_3
=-(\alpha_1+\alpha_2)$ and
\bea
PN^1_1=(1-x)^{(1-x)/x}\to \left(\frac{3}{7}\right)^{3/4}\approx 0.52968468 
\eea
which is very close to the ratio
${\rm Im}D^{12}_{3/2,1/2}/{\rm Im}D^{12}_{1/2,1/2}\approx 0.52967778$
evaluated using the last line of the tables ($M=1792$).

These comparisons of the bosonic overlap to the string
bit model for large finite $M$ give strong support to the use of
bosonization (with due care with phase conventions)
to calculate the continuum limit of spin chain models,
and, in particular, of  string bit models. A more systematic analysis of
the use of bosonization to calculate the 3 AP string vertex will be the
subject of another paper.

\vskip24pt
\begin{center}
  \begin{Large}
    {\bf Appendices}
\end{Large}
\end{center}

\appendix
      \section{Matrix elements}
      We quote a generally useful formula for matrix elements of the sort required in the overlap
      calculations:
      \bea
      &&\hskip-.5in
\bra{0}\exp\left\{\frac{1}{2}f^TAf+\alpha^T f\right\}
      \exp\left\{\frac{1}{2}f^{\dagger T}Bf^\dagger+\beta^T f^\dagger\right\}
      \ket{0}
      ={\det}^{1/2}(I+BA)\nonumber\\
      &&\times\exp\left\{
      \frac{1}{2}\alpha^TB(I+AB)^{-1}\alpha + \frac{1}{2}\beta^T(I+AB)^{-1}A
      \beta +\beta^T(I+AB)^{-1}\alpha\right\}
       \eea
       For example, this formula enables the normalization of the ground state
       of $h_{AP}$ when it is expressed in terms of the eigenstates of
       $h_P$, or vice versa.
      \section{Overlap Matrices}
      The transition amplitude between eigenstates of two different
      Hamiltonians can be determined by the relation between eigenoperators
      of the corresponding Hamiltonian. Call the lowering operators for
      $H_1$ and $H_2$  $f^1_k$ and $f^2_k$ respectively. Then define the
      matrices $C$ and $S$ by
      \bea
      f^2_k&=& C_{kl}f^1_l+S_{kl}f^{1\dagger}_l
      \eea
      To determine the ground state of $H_2$ in terms of the eigenstates of
      $H_1$, we need to solve the equation
      \bea
      CD=-S
      \eea
      where $D$ enters the ground state ket of $H_2$ as
      \bea
      \ket{G2}&=&\exp\left\{\frac{1}{2}f^{1\dagger}_kD_{kl}f^{1\dagger}_l
      \right\}\ket{G1}
      \eea
      We first quote the $C$ and $S$ matrices for two periodic strings ($H_1$)
      transitioning to one periodic string ($H_2$) obtained in
      \cite{thornprotobits}.
      \bea
C_{m0}&=&-\frac{1}{\sqrt{LK}}\frac{1-e^{-2\pi imL/M}}{1-e^{2i\pi m/M}}\cos\left(\frac{m\pi}{2M}-\frac{\pi}{4}\right)\nonumber\\
C_{mn1}&=&-\frac{1}{\sqrt{ML}}\frac{1-e^{-2\pi imL/M}}{1-e^{-2i\pi (n/L-m/M)}}\cos\left(\frac{n\pi}{2L}-\frac{m\pi}{2M}\right)\nonumber\\
C_{mn2}&=&\frac{1}{\sqrt{MK}}
\frac{1-e^{-2\pi imL/M}}{1-e^{-2i\pi (n/K-m/M)}}\cos\left(\frac{n\pi}{2K}
-\frac{m\pi}{2M}\right)\\
S_{m0}&=&-\frac{1}{\sqrt{LK}}\frac{1-e^{-2\pi imL/M}}{1-e^{2i\pi m/M}}\cos\left(\frac{m\pi}{2M}+\frac{\pi}{4}\right)\nonumber\\
S_{mn1}&=&-\frac{1}{\sqrt{ML}}\frac{1-e^{-2\pi imL/M}}{1-e^{2i\pi (n/L+m/M)}}
\cos\left(\frac{n\pi}{2L}+\frac{m\pi}{2M}\right)\nonumber\\
S_{mn2}&=&\frac{1}{\sqrt{MK}}
\frac{1-e^{-2\pi imL/M}}{1-e^{2i\pi (n/K+m/M)}}\cos\left(\frac{n\pi}{2K}
+\frac{m\pi}{2M}\right).\eea
The continuum limit of these matrices was given in \cite{thornprotobits}.

Replacing the two periodic strings with two antiperiodic strings leads to the
modifications
\bea
C_{0r1}&=&-\frac{1}{\sqrt{MK}}\frac{2}{1-e^{-2i\pi r/K}}\cos\left(
  \frac{r\pi}{2K}-\frac{\pi}{4}\right)\nonumber\\
C_{mr1}&=&-\frac{1}{\sqrt{MK}}
\frac{1+e^{-2\pi imK/M}}{1-e^{2i\pi (m/M-r/K)}}\cos\left(\frac{m\pi}{2M}
  -\frac{r\pi}{2K}\right)\nonumber\\
C_{0r2}&=&-\frac{1}{\sqrt{ML}}\frac{2}{1-e^{-2i\pi r/L}}\cos\left(
  \frac{r\pi}{2L}-\frac{\pi}{4}\right)\nonumber\\
C_{mr2}&=&-\frac{1}{\sqrt{ML}}
\frac{1+e^{-2\pi imK/M}}{1-e^{2i\pi (m/M-r/L)}}\cos\left(\frac{m\pi}{2M}
-\frac{r\pi}{2L}\right)\\
S_{0r1}&=&-\frac{1}{\sqrt{MK}}\frac{2}{1-e^{2i\pi r/K}}\cos\left(
  \frac{r\pi}{2K}+\frac{\pi}{4}\right)\nonumber\\
S_{mr1}&=&-\frac{1}{\sqrt{MK}}
\frac{1+e^{-2\pi imK/M}}{1-e^{2i\pi (m/M+r/K)}}\cos\left(\frac{m\pi}{2M}
  +\frac{r\pi}{2K}\right)\nonumber\\
S_{0r2}&=&-\frac{1}{\sqrt{ML}}\frac{2}{1-e^{2i\pi r/L}}\cos\left(
  \frac{r\pi}{2L}+\frac{\pi}{4}\right)\nonumber\\
S_{mr2}&=&-\frac{1}{\sqrt{ML}}
\frac{1+e^{-2\pi imK/M}}{1-e^{2i\pi (m/M+r/L)}}\cos\left(\frac{m\pi}{2M}
  +\frac{r\pi}{2L}\right).
\eea
The continuum limit of these matrices is $K,L\to\infty$ with $K/M=x$ with
$0<x<1$ fixed and $m$ or $M-m$, $r1$ or $K-r1$, and $r2$ or $L-r2$
finite.
It is convenient to remove some overall factors from $C$ and $S$. Define
\bea
C_{mr}&=&\frac{1+e^{-2\pi i xm}}{2\pi i}c_{mr},\qquad
  S_{mr}=\frac{1+e^{-2\pi i xm}}{2\pi i}s_{mr}
\eea
We quote the continuum limit in terms of the $c$ and $s$ matrices.
There are eight distinct cases:

1. $m$, $r1$, $r2$ finite
\bea
c_{0r1}&\to&-\frac{\sqrt{x}}{r\sqrt{2}}
\qquad c_{0r2}\to-\frac{\sqrt{1-x}}{r\sqrt{2}}\nonumber\\
c_{mr1}&\to&\frac{1}{m\sqrt{x}-r/\sqrt{x}}
\qquad c_{mr2}\to\frac{1}{m\sqrt{1-x}-r/\sqrt{1-x}}\\
s_{0r1}&\to&\frac{\sqrt{x}}{r\sqrt{2}}
\qquad s_{0r2}\to\frac{\sqrt{1-x}}{r\sqrt{2}}\nonumber\\
s_{mr1}&\to&\frac{1}{m\sqrt{x}+r/\sqrt{x}}
\qquad s_{mr2}\to\frac{1}{m\sqrt{1-x}+r/\sqrt{1-x}}
\eea

2. $m^\prime=M-m\neq0$, $r1$, $r2$ finite
\bea
c_{mr1}\sim O(M^{-1})\to0,\qquad c_{mr2}\sim O(M^{-1})\to0\\
    s_{mr1}\sim O(M^{-1})\to0,\qquad s_{mr2}\sim O(K^{-1})\to0
        \eea

3. $m$, $r^\prime1=K-r1$, $r2$ finite
\bea
c_{0r1}&\to& \frac{\sqrt{x}}{r^\prime\sqrt{2}},\qquad
  c_{0r2}\to- \frac{\sqrt{1-x}}{r\sqrt{2}}\nonumber\\
   c_{mr1}&\to&0,\qquad c_{mr_2}\to \frac{1}{m\sqrt{1-x}-r/\sqrt{1-x}}\\
   s_{0r_1}&\to&\frac{\sqrt{x}}{r^\prime\sqrt{2}},\qquad
   s_{0r2}\to \frac{\sqrt{1-x}}{r\sqrt{2}}\nonumber\\
   s_{mr1}&\to&0,\qquad s_{mr_2}\to \frac{1}{m\sqrt{1-x}+r/\sqrt{1-x}}
   \eea

4. $m^\prime$, $r^\prime1$, $r2$ finite
\bea
c_{mr1}&\to&-\frac{1}{m^\prime\sqrt{x}-r^\prime/\sqrt{x}},\qquad
c_{mr2}\to0\\
s_{mr1}&\to&\frac{1}{m^\prime\sqrt{x}+r^\prime/\sqrt{x}},\qquad
s_{mr2}\to0
\eea

5. $m$, $r1$, $r^\prime2$ finite
\bea
c_{0r1}&\to&-\frac{\sqrt{x}}{r\sqrt{2}}
\qquad c_{0r2}\to\frac{\sqrt{1-x}}{r^\prime\sqrt{2}}\nonumber\\
c_{mr1}&\to&\frac{1}{m\sqrt{x}-r/\sqrt{x}}
\qquad c_{mr2}\to0\\
s_{0r1}&\to&\frac{\sqrt{x}}{r\sqrt{2}}
\qquad s_{0r2}\to-\frac{\sqrt{1-x}}{r^\prime\sqrt{2}}\nonumber\\
s_{mr1}&\to&\frac{1}{m\sqrt{x}+r/\sqrt{x}}
\qquad s_{mr2}\to0
\eea

6. $m^\prime$, $r1$, $r^\prime2$ finite
\bea
c_{mr1}&\to&0,
\qquad c_{mr2}\to-\frac{1}{m^\prime\sqrt{1-x}-r^\prime/\sqrt{1-x}}\\
s_{mr1}&\to&0
\qquad s_{mr2}\to\frac{1}{m^\prime\sqrt{1-x}+r^\prime/\sqrt{1-x}}
\eea

7. $m$, $r^\prime1$, $r^\prime2$ finite
\bea
c_{0r1}&\to& \frac{\sqrt{x}}{r^\prime\sqrt{2}},\qquad
  c_{0r2}\to\frac{\sqrt{1-x}}{r\sqrt{2}}\nonumber\\
   c_{mr1}&\to&0,\qquad c_{mr_2}\to0\\
   s_{0r_1}&\to&\frac{\sqrt{x}}{r^\prime\sqrt{2}},\qquad
   s_{0r2}\to -\frac{\sqrt{1-x}}{r\sqrt{2}}\nonumber\\
   s_{mr1}&\to&0,\qquad s_{mr_2}\to 0
   \eea

8. $m^\prime$, $r^\prime1$, $r^\prime2$ finite
\bea
c_{mr1}&\to&-\frac{1}{m^\prime\sqrt{x}-r^\prime/\sqrt{x}}
\qquad c_{mr2}\to-\frac{1}{m^\prime\sqrt{1-x}-r^\prime/\sqrt{1-x}}\\
s_{mr1}&\to&\frac{1}{m^\prime\sqrt{x}+r^\prime/\sqrt{x}}
\qquad s_{mr2}\to\frac{1}{m^\prime\sqrt{1-x}+r^\prime/\sqrt{1-x}}
\eea

\section{The continuum limit of $D$ for 
  2 antiperiodic strings $\to$ 1 periodic string}
In \cite{thornprotobits} we obtained the continuum limit of $D$ for
the case that all strings are periodic using Goldstone's method
\cite{goldstone}, To do the analogous analysis for this case, we
work with a slightly modified meromorphic function
\bea
g(z)&=&\frac{\Gamma(1/2+xz)\Gamma(1/2+(1-x)z)}{\Gamma(1+z)\sqrt{x(1-x)}}
e^{\xi z}\nonumber\\
\xi&=& -x\ln x -(1-x)\ln(1-x) .
\eea
It has zeroes at negative integers and poles at $z=-(1/2+n)/x$ and
at $z=-(1/2+n)/(1-x)$ with $n=0,1,2,\cdots$:
\bea
g(z)&\sim&\frac{1}{z+r/x}\frac{1}{rg(r/x)},\qquad
  g(z)\sim\frac{1}{z+r/(1-x)}\frac{1}{rg(r/(1-x))}
  \eea
  respectively.
Its large $z$ behavior is
\bea
g(z)&\sim& \sqrt{2\pi} z^{-1/2},\qquad |\arg(z)|<\pi
\eea
For $m$ a positive integer, $g(z)/(z+m)$ has the same poles as $g(z)$,
since $g(-m)=0$. Thus we can expand
\bea
\frac{g(z)}{z+m}&=&\sum_r \left[
  \frac{1}{z+r/x}\frac{1}{rg(r/x)}\frac{1}{m-r/x}\right.\nonumber\\
&& \left.+\frac{1}{z+r/(1-x)}\frac{1}{rg(r/(1-x))}\frac{1}{m-r/(1-x)}\right],
\qquad m=1,2,\cdots.
\eea
Another identity follows by expanding
\bea
\frac{zg(z)}{z+m}&=&\sum_r \left[
  \frac{-r/x}{z+r/x}\frac{1}{rg(r/x)}\frac{1}{m-r/x}\right.\nonumber\\
&& \left.+\frac{-r/(1-x)}{z+r/(1-x)}\frac{1}{rg(r/(1-x))}\frac{1}{m-r/(1-x)}
\right],\qquad m=0,1,2,\cdots.
\eea
Note that this second identity is valid for $m=0$, which will be
important for
a complete determination of $D$. Also note that the right sides of
both identities are symmetric under $x\to 1-x$.

Setting $z=s/x$ or $z=s/(1-x)$ turns these identities into equations similar
to those to be solved. And comparing them in various cases, allows
one to read off explicit expressions for the matrix elements of $D$:
\bea
\frac{1}{m+s/x}&=&\sum_r \left[
  \frac{1}{s/x+r/x}\frac{1}{rg(r/x)g(s/x)}\frac{1}{m-r/x}\right.\nonumber\\
&& \left.+\frac{1}{s/x+r/(1-x)}\frac{1}{rg(r/(1-x))g(s/x)}
  \frac{1}{m-r/(1-x)}\right]
\\
\frac{1}{m+s/x}&=&\sum_r \left[
  \frac{-r/x}{s/x+r/x}\frac{1}{rg(r/x)(s/x)g(s/x)}
  \frac{1}{m-r/x}\right.\nonumber\\
&& \left.+\frac{-r/(1-x)}{s/x+r/(1-x)}\frac{1}{rg(r/(1-x))(s/x)g(s/x)}
  \frac{1}{m-r/(1-x)}
  \right]\\
\frac{1}{m+s/(1-x)}&=&\sum_r \left[
  \frac{1}{s/(1-x)+r/x}\frac{1}{rg(r/x)g(s/(1-x))}\frac{1}{m-r/x}\right.
\nonumber\\
&&\hskip-1cm \left.+\frac{1}{s/(1-x)+r/(1-x)}\frac{1}{rg(r/(1-x))g(s/(1-x))}
  \frac{1}{m-r/(1-x)}\right]
\\
\frac{1}{m+s/(1-x)}&=&\sum_r \left[
  \frac{-r/x}{s/(1-x)+r/x}\frac{1}{rg(r/x)(s/(1-x))g(s/(1-x))}
  \frac{1}{m-r/x}\right.\nonumber\\
&& \hskip-3cm\left.+\frac{-r/(1-x)}{s/(1-x)+r/(1-x)}
  \frac{1}{rg(r/(1-x))(s/(1-x))g(s/(1-x))}
  \frac{1}{m-r/(1-x)}
  \right]
  \eea
  Again the second pair of equations is the first pair with $x\to1-x$.
The sum of the two identities enables the construction of antisymmetric
solutions for $D_{rs}$ or $D_{r^\prime s^\prime}$. (As before we use a prime
to distinguish indices close to their upper limit.)
\bea
\frac{1}{m+s/x}&=&\frac{1}{2}\sum_r \left[
  \frac{s-r}{s/x+r/x}\frac{1}{rg(r/x)sg(s/x)}\frac{1}{m-r/x}\right.\nonumber\\
&& \left.+\frac{s-xr/(1-x)}{s/x+r/(1-x)}\frac{1}{rg(r/(1-x))sg(s/x)}
  \frac{1}{m-r/(1-x)}\right]\\
\frac{1}{m+s/(1-x)}&=&\frac{1}{2}\sum_r \left[
  \frac{s-(1-x)r/x}{s/(1-x)+r/x}\frac{1}{rg(r/x)sg(s/(1-x))}
  \frac{1}{m-r/x}\right.
\nonumber\\
&& \hskip-3cm\left.+\frac{s-r}{s/(1-x)+r/(1-x)}\frac{1}{rg(r/(1-x))sg(s/(1-x))}
  \frac{1}{m-r/(1-x)}\right]
\eea
On the other hand the difference of the two identities is useful for the
construction of solutions for $D_{rs^\prime}$ or $D_{r^\prime s}$:
\bea
0&=&\sum_r \left[
  \frac{x}{rg(r/x)sg(s/x)}\frac{1}{m-r/x}\right.\nonumber\\
&&\hskip+1cm \left.+\frac{x}{rg(r/(1-x))sg(s/x)}
  \frac{1}{m-r/(1-x)}\right]
\\
0&=&\sum_r \left[
  \frac{1-x}{rg(r/x)sg(s/(1-x))}\frac{1}{m-r/x}\right.
\nonumber\\
&&\hskip+1cm \left.+\frac{1-x}{rg(r/(1-x))sg(s/(1-x))}
  \frac{1}{m-r/(1-x)}\right]\eea
Notice that the $r$ and $s$ dependence factorizes, which shows that
the second equation is an $s$ dependent factor times the first. 
The content of either is therefore
\bea
0&=&\sum_r \left[
  \frac{1}{rg(r/x)}\frac{1}{m-r/x}+\frac{1}{rg(r/(1-x))}
  \frac{1}{m-r/(1-x)}\right],\qquad m=1.2.\cdots
\eea
Finally we recall that the $m=0$ case of the second identity is also valid
and can be rewritten:
\bea
g(s/x)&=&\sum_r \left[\frac{1}{s/x+r/x}\frac{1}{rg(r/x)}
  +\frac{1}{s/x+r/(1-x)}\frac{1}{rg(r/(1-x))}\right]\\
g(s/(1-x))&=&\sum_r \left[
  \frac{1}{s/(1-x)+r/x}\frac{1}{rg(r/x)}
\right.\nonumber\\&& \hskip1cm\left.
  +\frac{1}{s/(1-x)+r/(1-x)}
  \frac{1}{rg(r/(1-x))}
  \right]
  \eea
  Once again, the second equation is simply the first with the substitution
  $x\to1-x$.

  We next write out the equation $cD=-s$ in the continuum limit, employing our
  convention about primed and unprimed indices:. There are 12 types of matrix
  elements of $c$ and $s$: the first index can be $0,m\neq0, m^\prime$ and the
  second can be $s1,s^\prime1,s2,s^\prime2$. Matrix elements of $c,s$ that
  link primed indices and nonzero unprimed indices are zero.
Let's begin with the $ms1$ matrix element
  \bea
  -s_{ms1}=-\frac{1/\sqrt{x}}{m+r/x}&=&c_{mr1}D_{r1s1}+c_{mr2}D_{r2s1}
  \nonumber\\
  &=&\frac{1/\sqrt{x}}{m-r/x}D_{r1s1}+\frac{1/\sqrt{1-x}}{m-r/(1-x)}D_{r3s1}
  \eea
  Comparing to the first identity we read off
  \bea
  D_{r1s1}&=&-\frac{1}{2}\frac{s-r}{s/x+r/x}\frac{1}{rg(r/x)sg(s/x)}\\
  D_{r2s1}&=&-\frac{1}{2}\sqrt{\frac{1-x}{x}}\frac{s-xr/(1-x)}{s/x+r/(1-x)}
  \frac{1}{rg(r/(1-x))sg(s/x)}
  \eea
A parallel analysis of the $ms2$ matrix element
  \bea
  -s_{ms2}=-\frac{1}{m\sqrt{1-x}+r/\sqrt{1-x}}
  &=&c_{mr1}D_{r1s2}+c_{mr2}D_{r2s2}\nonumber\\
  &=&\frac{1/\sqrt{x}}{m-r/x}D_{r1s2}+\frac{1/\sqrt{1-x}}{m-r/(1-x)}D_{r2s2}
  \eea
  yields
  \bea
  D_{r1s2}&=&-\frac{1}{2}\sqrt{\frac{x}{1-x}}
  \frac{s-(1-x)r/x}{s/(1-x)+r/x}\frac{1}{rg(r/x)sg(s/(1-x))}\\
  D_{r2s2}&=&-\frac{1}{2}\frac{s-r}{s/(1-x)+r/(1-x)}
  \frac{1}{rg(r/(1-x))sg(s/(1-x))}
  \eea
  Note that $D_{r1s1}$ and $D_{r2s2}$ are explicitly antisymmetric,
  and they both go into each other on the substitution $x\to 1-x$.
  for the mixed indices $D_{r1s2}=-D_{s2r1}$ by inspecting the results of
  the two calculations.

  The elements so far obtained have all indices a finite
  distance from $0$. When all indices are a finite distance
  from their upper limits, we can obtain the $D$ elements by noting that
  in the latter case $c$ and $s$ can be obtained from the former case by
  priming all indices and multiplying the $c$ matrix elements by $-1$.
  It follows that corresponding $D$ elements are the negatives of those
  in the unprimed case.
  
 \bea
 D_{r^\prime1s^\prime1}&=&\frac{1}{2}
 \frac{s^\prime-r^\prime}{s^\prime/x+r^\prime/x}
 \frac{1}{r^\prime g(r^\prime/x)s^\prime g(s^\prime/x)}\\
 D_{r^\prime2s^\prime1}&=&\frac{1}{2}\sqrt{\frac{1-x}{x}}
 \frac{s^\prime-xr^\prime/(1-x)}{s^\prime/x+r^\prime/(1-x)}
  \frac{1}{r^\prime g(r^\prime/(1-x))s^\prime g(s^\prime/x)}\\
  D_{r^\prime1s^\prime2}&=&\frac{1}{2}\sqrt{\frac{x}{1-x}}
  \frac{s^\prime-(1-x)r^\prime/x}{s^\prime/(1-x)+r^\prime/x}
  \frac{1}{r^\prime g(r^\prime/x)s^\prime g(s^\prime/(1-x))}\\
  D_{r^\prime2s^\prime2}&=&\frac{1}{2}
  \frac{s^\prime-r^\prime}{s^\prime/(1-x)+r^\prime/(1-x)}
  \frac{1}{r^\prime g(r^\prime/(1-x))s^\prime g(s^\prime/(1-x))}
  \eea
  It remains to obtain the mixed $D$ elements with one index unprimed and
  the other primed.

  We begin with the equation (and the one with $s^\prime2$ in place of
  $s^\prime1$.)
  \bea
  s_{ms^\prime1}=0&=& c_{mr1}D_{r1s^\prime 1}+c_{mr2}D_{r2s^\prime 1}
  \nonumber\\
&=&  \sum_r\left[\frac{1/\sqrt{x}}{m-r/x}D_{r1s^\prime 1}
  +\frac{1/\sqrt{1-x}}{m-r/(1-x)}D_{r2s^\prime 1}\right]
\eea
Referring to the homogeneous identity we read off
\bea
D_{r1s^\prime 1}&=&\kappa_1(s^\prime)\frac{\sqrt{x}}{rg(r/x)}\\
D_{r2s^\prime 1}&=&\kappa_1(s^\prime)\frac{\sqrt{1-x}}{rg(r/(1-x))}
\eea
$\kappa(s^\prime)$ is undetermined because the equation is homogeneous.
Substituting $s^\prime2$ for $s^\prime1$ in the equation doesn't change the
$r$ dependence but the $\kappa$ may be different:
\bea
D_{r1s^\prime 2}&=&\kappa_2(s^\prime)\frac{\sqrt{x}}{rg(r/x)}\\
D_{r2s^\prime 2}&=&\kappa_2(s^\prime)\frac{\sqrt{1-x}}{rg(r/(1-x))}
\eea
The equation
 \bea
 s_{m^\prime s1}=0&=& c_{m^\prime r^\prime1}D_{r^\prime1s1}
 +c_{m^\prime r^\prime2}D_{r^\prime2s1}
  \nonumber\\
&=&  \sum_r\left[\frac{1/\sqrt{x}}{m-r/x}D_{r1s^\prime 1}
  +\frac{1/\sqrt{1-x}}{m-r/(1-x)}D_{r2s^\prime 1}\right]
\eea
and the one with $s1\to s2$ constrain the mixed $D$ elements with the
first index primed:
\bea
D_{r^\prime 1s 1}&=&\kappa^\prime_1(s)\frac{\sqrt{x}}{r^\prime g(r^\prime/x)}\\
D_{r^\prime2s 1}&=&\kappa^\prime_1(s)
\frac{\sqrt{1-x}}{r^\prime g(r^\prime/(1-x))}\\
D_{r^\prime 1s 2}&=&\kappa^\prime_2(s)\frac{\sqrt{x}}{r^\prime g(r^\prime/x)}\\
D_{r^\prime2s 2}&=&\kappa^\prime_2(s)
+\frac{\sqrt{1-x}}{r^\prime g(r^\prime/(1-x))}
\eea
Again the $\kappa$'s can be different in each case. To determine them we turn
to the $m=0$ equations.
\bea
  -s_{0s1}&=&c_{0r1}D_{r1s1}+c_{0r2}D_{r2s1}+c_{0r^\prime1}D_{r^\prime1s1}
  +c_{0r^\prime2}D_{r^\prime2s1}\\
  -s_{0s2}&=&c_{0r1}D_{r1s2}+c_{0r2}D_{r2s2}+c_{0r^\prime1}D_{r^\prime1s2}
  +c_{0r^\prime2}D_{r^\prime2s2}
   \eea
   together with the two equations with $s$ primed.
   The first, in the continuum limit, reads
   \bea
   -\frac{\sqrt{x}}{s\sqrt{2}}&=&-\frac{\sqrt{x}}{r\sqrt{2}}D_{r1s1}
   -\frac{\sqrt{1-x}}{r\sqrt{2}}D_{r2s1}
   +\frac{\sqrt{x}}{r^\prime\sqrt{2}}D_{r^\prime1s1}
   +\frac{\sqrt{1-x}}{r^\prime\sqrt{2}}D_{r^\prime2s1}
   \eea
   The sums in the first two terms can be simplified by writing the numerator
   in the $D$'s as a linear combination of $s$ and the denominator in the $D$:
   $s-r=s+r-2r$ for $D_{r1s1}$ and $s+xr/(1-x)-2xr/(1-x)$. The contribution of
   the first term cancels the denominator and the second enters
   as a term in the zero mode identity.
   
   \bea
  c_{0r1}D_{r1s1}+c_{0r2}D_{r2s1}&=&\frac{\sqrt{x}}{2\sqrt{2}sg(s/x)}
  \sum_r\left[\frac{x}{r^2g(r/x)}+\frac{(1-x)}{r^2g(r/(1-x))}\right]
  \nonumber\\
  &&\hskip-2.5cm -\frac{\sqrt{x}}{\sqrt{2}sg(s/x)}
  \sum_r\left[\frac{x}{s+r}
   \frac{1}{rg(r/x)}+\frac{1}{s/x+r/(1-x)}\frac{1}{rg(r/(1-x))}\right]
 \nonumber\\
 &=&\frac{\sqrt{x}}{2\sqrt{2}sg(s/x)}
 \sum_r\left[\frac{x}{r^2g(r/x)}+\frac{(1-x)}{r^2g(r/(1-x))}\right]
 -\frac{\sqrt{x}}{\sqrt{2}s}
\eea
where we used the zero mode identity to get the last line. Plugging
this in, the equation becomes
\bea
 0&=&\frac{\sqrt{x}}{2\sqrt{2}sg(s/x)}
 \sum_r\left[\frac{x}{r^2g(r/x)}+\frac{(1-x)}{r^2g(r/(1-x))}\right]
 \nonumber\\
 &&+\kappa^\prime_1(s)\left[\frac{{x}}{r^{\prime2}g(r^\prime/x)\sqrt{2}}
   +\frac{1-x}{r^{\prime2}g(r^\prime/(1-x)\sqrt{2}}\right]\\
 \kappa^\prime_1(s)&=&-\frac{\sqrt{x}}{2sg(s/x)}
 \eea
 Applying the same procedure to the remaining 3 zero mode equations determines
 the remaining $\kappa$'s:
 \bea
 \kappa^\prime_2(s)&=&-\frac{\sqrt{1-x}}{2sg(s/(1-x))}\nonumber\\
 \kappa_1(s^\prime)&=&\frac{\sqrt{x}}{2sg(s^\prime/x)}\nonumber\\
 \kappa_2(s^\prime)&=&\frac{\sqrt{1-x}}{2sg(s^\prime/(1-x))}\nonumber
 \eea
We have thus fully determined the mixed matrix elements of $D$:
 \bea
 D_{r^\prime1 s1}&=&-\frac{{x}}{2r^\prime g(r^\prime/x)sg(s/x)}\\
 D_{r^\prime2 s1}&=&-\frac{\sqrt{x(1-x)}}{2r^\prime g(r^\prime/(1-x))sg(s/x)}\\
 D_{r^\prime1 s2}&=&-\frac{\sqrt{x(1-x)}}{2r^\prime g(r^\prime/x)sg(s/(1-x))}\\
 D_{r^\prime2 s2}&=&-\frac{1-x}{2r^\prime g(r^\prime/(1-x))sg(s/(1-x))}\\
 D_{r1 s^\prime1}&=&\frac{{x}}{2r g(r/x)s^\prime g(s^\prime/x)}\\
 D_{r2 s^\prime1}&=&\frac{\sqrt{x(1-x)}}{2r g(r/(1-x))s^\prime g(s^\prime/x)}\\
 D_{r1 s^\prime2}&=&\frac{\sqrt{x(1-x)}}{2r g(r/x)s^\prime g(s^\prime/x)}\\
 D_{r2 s^\prime2}&=&\frac{1-x}{2r g(r/(1-x))s^\prime g(s^\prime/(1-x))}
 \eea
 One can easily verify the antisymmetry of these matrix elements
 \bea
 D_{r^\prime1 s1}&=&-D_{s1 r^\prime1},
 \qquad D_{r^\prime2 s1}=- D_{s1 r^\prime2}\nonumber\\
 D_{r^\prime1 s2}&=&-D_{s2 r^\prime1},
 \qquad   D_{r^\prime2 s2}=-D_{s2 r^\prime2}\nonumber
 \eea
 \vskip12pt
 \noindent {\large  {\bf Acknowledgements}}: I thank the high energy theory group
and the Department of Physics at the
University of Florida for their hospitality and support.

\end{document}